\documentclass[reprint,amsmath,amssymb,aps]{revtex4-2}
\usepackage{graphicx}
\usepackage{dcolumn}
\usepackage{bm}
\usepackage[utf8]{inputenc}
\usepackage[T1]{fontenc}
\usepackage{booktabs, array, mathptmx, float, tabularx, booktabs, lipsum, amsmath,multirow}
\usepackage{siunitx, xcolor}
\usepackage[version=4]{mhchem}

\begin{document}

\title{Two-dimensional superconductors with intrinsic $p$-wave pairing or nontrivial band topology}
\author{Wei Qin$^1$$^,$$^2$}
\author{Jiaqing Gao$^1$}
\author{Ping Cui$^1$$^*$}
\author{Zhenyu Zhang$^1$$^\dagger$}
\address{$^1$International Center for Quantum Design of Functional Materials (ICQD),\\ Hefei National Research Center for Physical Sciences at Microscale,\\ University of Science and Technology of China, Hefei, Anhui 230026, China}
\address{$^2$Department of Physics, The University of Texas at Austin, Austin, Texas 78712, USA}
\collaboration{$^*$email: cuipg@ustc.edu.cn; $^\dagger$email: zhangzy@ustc.edu.cn}

\begin{abstract}
Over the past fifteen years, tremendous efforts have been devoted to realizing topological superconductivity in realistic materials and systems, predominately propelled by their promising application potentials in fault-tolerant quantum information processing. In this article, we attempt to give an overview on some of the main developments in this field, focusing in particular on two-dimensional crystalline superconductors that possess either intrinsic $p$-wave pairing or nontrivial band topology. We first classify the three different conceptual schemes to achieve topological superconductor (TSC), enabled by real-space superconducting proximity effect, reciprocal-space superconducting proximity effect, and intrinsic TSC. Whereas the first scheme has so far been most extensively explored, the subtle difference between the other two remains to be fully substantiated. We then move on to candidate intrinsic or $p$-wave superconductors, including Sr$_2$RuO$_4$, UTe$_2$, Pb$_3$Bi, and graphene-based systems. For TSC systems that rely on proximity effects, the emphases are mainly on the coexistence of superconductivity and nontrivial band topology, as exemplified by transition metal dichalcogenides, cobalt pnictides, and stanene, all in monolayer or few-layer regime. The review completes with discussions on the three dominant tuning schemes of strain, gating, and ferroelectricity in acquiring one or both essential ingredients of the TSC, and optimizations of such tuning capabilities may prove to be decisive in our drive towards braiding of Majorana zero modes and demonstration of topological qubits.
\end{abstract}

\maketitle

\section{Introduction}\label{sec:intro}
The last two decades have witnessed the great success of the concept of topology applied to condensed matter physics, which not only enriches our knowledge of quantum phases of matter but also provides topological material-based unique technical applications. In mathematics, topology is used to classify geometric shapes, for example, two shapes are topologically inequivalent if they cannot be continuously deformed into each other. Similarly, if the wave function of a quantum many-body system cannot be adiabatically connected to its trivial atomic limit, the system is classified as topologically nontrivial and characterized by the emergence of robust boundary states \cite{Hasan:2010aa,Qi:2011aa}. The integer quantum Hall insulator is a paradigm example of topologically nontrivial systems \cite{Klitzing:1980aa}, where the precisely quantized Hall conductance has a topological origin characterized by the Thouless-Kohmoto-Nightingale-de Nijs number or Chern number \cite{Thouless:1982aa}. In 2005, Kane and Mele found that the spin-orbit coupling can covert graphene from a semimetal into a quantum spin Hall insulator \cite{Kane:2005aa}, a spinful version of Haldane's model \cite{Haldane:1988aa}. This discovery triggered the widespread explorations on topological insulators in condensed matter systems. To date, the topological concept has been generalized to various electronic systems, such as insulators \cite{Hasan:2010aa,Qi:2011aa,Chiu:2016aa}, metals/semimetals \cite{Burkov:2016aa,Bansil:2016aa,Armitage:2018aa}, and superconductors \cite{Qi:2011aa,Alicea:2012aa,Sato:2017aa}, where the presence of a quasiparticle band gap, direct,  indirect, or curved, is indispensable for defining the topology. The establishment of the database for topological materials \cite{Bradlyn:2017aa,Tang:2019aa,Vergniory:2019aa,Vergniory:2021aa} has further completed the band theory of solids. Beyond the electronic systems, topological states have been unveiled in photonic crystals \cite{Lu:2014aa,Ozawa:2019aa}, phononic systems \cite{Ma:2019aa,Liu:2020aa}, and non-Hermitian systems \cite{Yao:2018aa,Gong:2018aa,Bergholtz:2021aa}. Moreover, the concept of topology has been recently extended to higher orders \cite{Schindler:2018aa,Khalaf:2018aa,Kim:2020aa,Xie:2021aa}, keeping this field attractive.

In analogy to insulators, if the wave function of a superconductor cannot be adiabatically connected to the trivial Bose-Einstein condensate of Cooper pairs, the superconductor is classified as topologically nontrivial \cite{Qi:2011aa}. The corresponding boundary state of a topological superconductor (TSC) is a superposition of particle and hole states, and therefore can be utilized to realize Majorana fermion \cite{Qi:2011aa,Alicea:2012aa}. Majorana fermion is its own antiparticle named after Ettore Majorana who deduced the real solutions of the Dirac equation \cite{Majorana:2008aa}. In 2000, Read and Green demonstrated that a two-dimensional (2D) spinless chiral $p$-wave superconductor can harbor Majorana zero mode (MZM) around a quantized magnetic vortex and chiral Majorana edge modes at the boundaries \cite{Read:2000aa}. The 1D counterpart of a $p$-wave superconductor or the Kitaev chain was shown to support unpaired Majorana fermions at the two ends of the chain \cite{Kitaev:2001aa}. The operator for exchanging two MZMs bounded by magnetic vortices possesses a necessary form of irreducible unitary matrix \cite{Read:2000aa,Ivanov:2001aa}, implying MZMs are subject to non-Abelian statistics \cite{Nayak:2008aa}. Because an ordinary fermion can be formally decomposed into two Majorana fermions, a pair of well separated Majorana fermions can be utilized to characterize a quantum state. Such a non-local way of information storage has an exceptional advantage of being immune to local noises, making Majorana fermion an ideal building block for realizing fault-tolerant quantum computing \cite{Nayak:2008aa,Sarma:2015aa}. 

Motivated by the promising technique application of Majorana fermion, tremendous efforts have been devoted to realizing TSC in realistic material systems \cite{Qi:2011aa,Alicea:2012aa,Leijnse:2012aa,Beenakker:2013aa,Stanescu:2013aa,Elliott:2015aa,Sato:2017aa,Li:2019aa,sharma:2022aa}. Topological property of a superconductor is characterized by the phase winding of order parameter around its Fermi surface \cite{Qi:2011aa,Alicea:2012aa}, which suggests odd-parity spin-triplet superconductor is a promising ground for realizing TSC \cite{Ivanov:2001aa}. To date, there are rare materials showing signatures of spin-triplet pairing, and various artificial schemes have been proposed to realize TSC. Depending on the way of introducing nontrivial phase winding into the superconducting order parameter, recipes for realizing TSC can be classified into three distinct categories: (i) real-space superconducting proximity effect-induced TSC; (ii) reciprocal-space superconducting proximity effect-induced TSC; (iii) intrinsic TSC. 

The scheme of realizing TSC via real-space proximity effect is first proposed by Fu and Kane in 2008 \cite{Fu:2008aa}. In their proposal, a heterostructure consisting of a 3D topological insulator and a conventional $s$-wave superconductor is shown to be an effective chiral $p$-wave superconductor. This idea has been experimentally explored in the Bi$_2$Te$_3$/NbSe$_2$ heterostructure \cite{Xu:2014aa,Xu:2015aa,Sun:2016aa}, where signature of Majorana fermion was revealed in the spin-polarized scanning tunneling spectroscopy measured around a magnetic vortex \cite{Sun:2016aa}. This real-space proximity scheme has been generalized to heterostructure comprised of a 2D or 1D semiconductor with spin-orbit coupling proximity coupled to spin-singlet superconductors \cite{Sau:2010aa,Lutchyn:2010aa,Oreg:2010aa,Alicea:2011aa,Zhang:2013aa,Lutchyn:2018aa,Prada:2020aa}. 
For reciprocal-space proximity or band proximity effect-induced TSC, typical examples are doped 3D topological insulators \cite{Hor:2010aa,Sasaki:2011aa,Levy:2013aa,Liu:2015aa} and iron-based superconductors \cite{Xu:2016aa,Shi:2017aa,Zhang:2017aa,Zhang:2018ab,Liu:2018aa,Hao:2018aa,Zhang:2019aa,Kong:2019aa,Wang:2020aa,Li:2022aa}, where superconductivity arising from the bulk bands is proximity coupled to the topological surface states, mimicking effective 2D chiral $p$-wave superconductors. Intrinsic TSC means that superconductivity and nontrivial topology arise from the same electronic states of the same material \cite{Elliott:2015aa,Sato:2017aa,sharma:2022aa}. There are increasing numbers of materials showing signatures of intrinsic topological superconductivity, such as Sr$_2$RuO$_4$ \cite{Maeno:1994aa}, UTe$_2$ \cite{Jiao:2020aa}, Pb$_3$Bi \cite{Qin:2009aa}. In addition to the three primary schemes, second-order TSCs are also proposed in 3D and 2D systems \cite{Wang:2018ab,Wang:2018ac,Hsu:2018aa,Zhu:2019aa,Wu:2020aa}, which are characterized by Majorana hinge and corner modes, respectively. 

Since TSC is uniquely characterized by Majorana boundary states, the following measurements are frequently involved in experimental demonstration of TSC. (i) Detection of zero bias conductance peaks (ZBCPs). The MZM is expected to emerge around a magnetic vortex core in a TSC, giving rise to a conductance peak in the scanning tunneling spectroscopy at zero bias voltage \cite{Law:2009aa,Flensberg:2010aa,Xu:2015aa,Sun:2016aa}. Nevertheless, the emergence of ZBCP only serves as a preliminary screening in pursuing TSC because other effects, such as in-gap Caroli-de Gennes-Matricon states induced by material imperfection \cite{Caroli:1994aa}, can also result in a conductance peak in the scanning tunneling spectroscopy. Although the MZM-induced ZBCP has been theoretically shown to exhibit a quantized value of $2e^2/h$, the experimental detection of this quantum effect remains highly controversial with discouraging or encouraging reports \cite{Zhang:2018aa,Chen:2019aa,Zhu:2020aa}. (ii) Measurements of thermal conductivity and spin current. Due to its superconducting nature, Majorana boundary states are hard to be directly detected via conventional electric transport measurements. However, the thermally insulating bulk state of a TSC makes it possible to evidence Majorana boundary states via thermal transport \cite{Wang:2011aa,Nomura:2012aa,Shiozaki:2013aa}. In particular, the thermal Hall conductance of a TSC has been shown to exhibit a quantized value, manifesting as quantized thermal Hall effect \cite{Beenakker:2016aa}. Moreover, a spin current associated with heat current can be detected if the boundary states possess spin polarization \cite{Sato:2017aa}. (iii) Measurements of anomalous Josephson effect. The current-phase relationship in a Josephson junction that involves TSC has been proposed to exhibit anomalous behaviors \cite{Beenakker:2013aa,Sato:2017aa}. For example, a junction between two 1D TSCs that contain a pair of Majorana fermions leads to the "fractional" Josephson effect \cite{Kwon:2003aa,Fu:2009aa,Lutchyn:2010aa}, characterized by $4\pi$-periodicity in the supercurrent phase difference \cite{Kitaev:2001aa,Jiang:2011aa}. Such an exotic behavior has been recognized as an important signature of topological superconductivity \cite{Rokhinson:2012aa,Jose:2012aa,Lee:2014ab,Wang:2018ad}. (iv) Angle-resolved photoemission spectroscopy (ARPES). For the real-space/reciprocal-space proximity effect-induced TSC, the Dirac-cone-type surface/edge states are expected to be fully gapped, which can be directly observed via high resolution ARPES \cite{Wang:2012ab,Zhang:2018ab}. (v) Measurements of spin texture. The surface/edge current of an intrinsic spin-triplet TSC is one of its key characteristics \cite{Kallin:2016aa}.

In this review, we summarize recent progresses in pursuing intrinsic TSCs, focusing on 2D candidates and their normal-state band topology. Intrinsic TSC is commonly associated with unconventional pairing that is sensitive to the quality of the hosting material. Recent advances in fabrication techniques make it possible to realize highly crystalline 2D superconductors, providing necessary conditions for realizing and studying intrinsic topological superconductivity. In addition to topological properties, 2D superconductors exhibit many other intriguing properties, such as oscillation of critical temperature due to quantum size effect \cite{Guo:2004aa,Eom:2006aa,Qin:2009aa}, Berezinskii–Kosterlitz–Thouless transition \cite{Berezinskii:1971aa,Berezinskii:1972aa,Kosterlitz:1972aa}, tunable superconductor-metal-insulator quantum phase transition \cite{Reyren:2007aa,Caviglia:2008aa,Saito:2015aa}, and in-plane critical magnetic field far beyond Pauli limit \cite{Lu:2015aa,Saito:2016aa,Xi:2016aa}. These properties have been adequately reviewed in ref.~\cite{Saito:2016ab}. This review in together with ref.~\cite{Saito:2016ab} complete the properties of 2D superconductors. The rest of this review is organized as follows. Sec.~\ref{sec:proximity} briefly introduces the basic ideas and some representative examples of realizing TSC via real- and reciprocal-space superconducting proximity effects. Sec.~\ref{sec:intrinsic} reviews candidate materials that harbor intrinsic topological superconductivity. Sec.~\ref{sec:coexist} summarizes the recently discovered 2D materials where superconductivity and band topology coexist. Sec.~\ref{sec:tunable} discusses three dominant tuning schemes, including strain, gating, and ferroelectricity, in acquiring tunable TSC. Sec.~\ref{sec:conclusion} gives a conclusion and perspective on the application of TSC and Majorana fermion in quantum computation. 

\section{TSC induced by proximity effect }
\label{sec:proximity}
The topological property of a superconductor is characterized by the phase winding of superconducting order parameter around the Fermi surface \cite{Qi:2011aa,Alicea:2012aa}. 
Within mean-field theory, the superconducting order parameter is defined as
\begin{equation}
\Delta(\bm{k}) = -\sum_{\bm{k}'} V(\bm{k},\bm{k}') \langle c_{-\bm{k}'} c_{\bm{k}'}\rangle,
\label{eq:gapequation}
\end{equation}
where $V(\bm{k},\bm{k}')$ denotes the interaction matrix, $c_{\bm{k}}$ ($c^{\dagger}_{\bm{k}}$) is the electron annihilation (creation) operator, spin and other degrees of freedom are implied. It is obvious from Eq.~(\ref{eq:gapequation}) that both of the interaction matrix $V(\bm{k},\bm{k}')$ and the single-particle wave function give contributions to the phase of $\Delta(\bm{k})$. For a proximity effect-induced TSC, the pairing glue and phase winding of $\Delta(\bm{k})$ originate from different while proximity coupled electronic states. In this section, we briefly review some representative examples of proximity effect-induced TSC by further distinguishing the proximity effects into real-space and reciprocal-space ones.

\subsection{Real-space proximity effect-induced TSC} 
\label{sec:real-proximity}

\begin{figure}[H]
\centering
\includegraphics[scale=0.65]{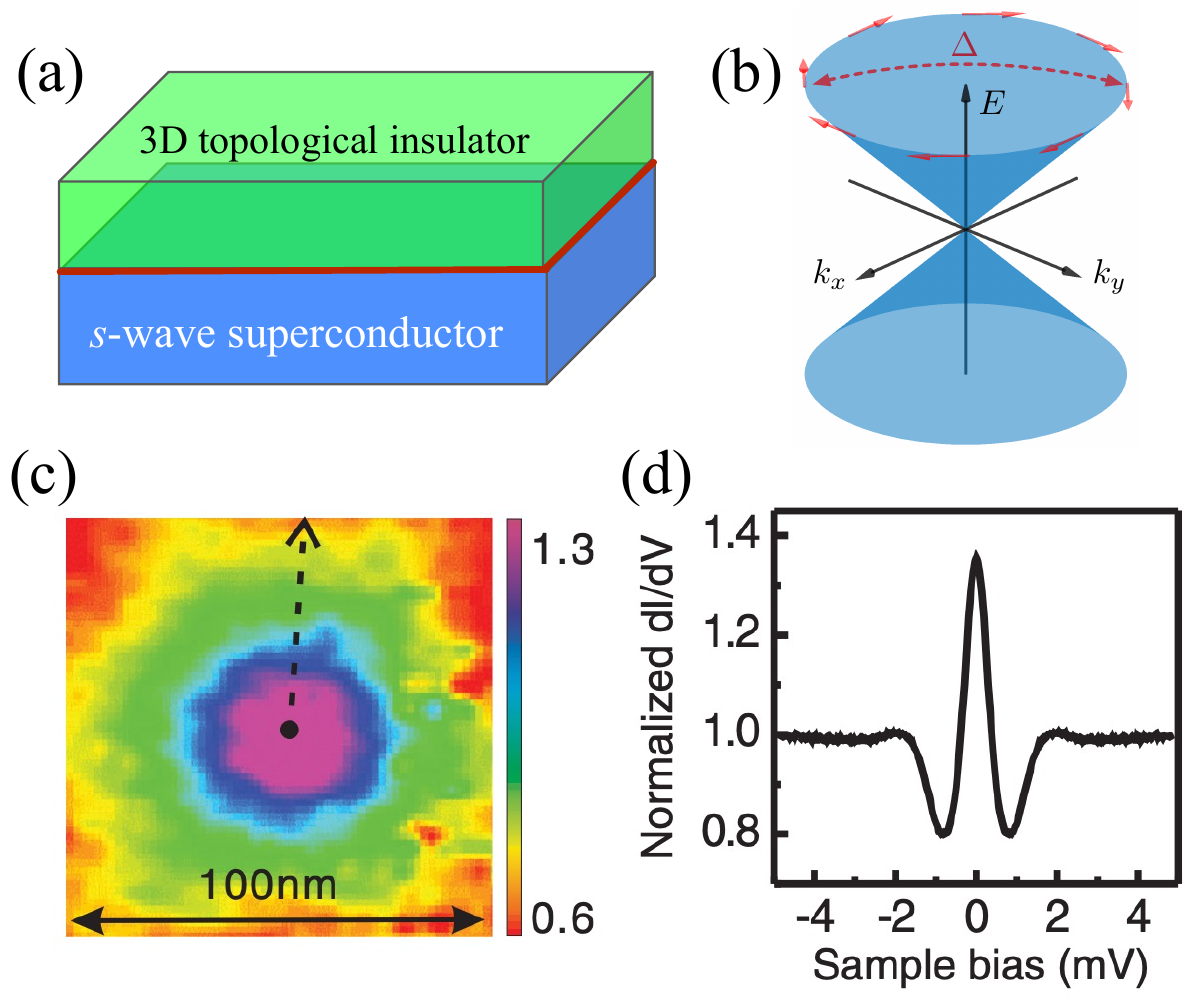}
\caption{(Color online) (a) Schematic of a 3D topological insulator/$s$-wave superconductor heterostructure. (b) Proximity effect-induced superconducting gap on the Dirac-type topological surface state. (c) A single magnetic vortex measured by zero-bias STS on the surface of Bi$_2$Te$_3$/NbSe$_2$ heterostructure. (d) Line cut along the direction marked in (c), where the zero-bias peak at the vortex center is a signature of a MZM. (c) and (d) are reprinted from ref.~\cite{Xu:2015aa} with permission from American Physical Society.} 
\label{fig:figure1}
\end{figure}

In 2008, Fu and Kane showed that an effective chiral $p$-wave superconductor can be realized at the interface of a 3D topological insulator and an $s$-wave superconductor \cite{Fu:2008aa}, as schematically depicted in Fig.~\ref{fig:figure1}(a). Superconductivity is induced in the topological surface states via proximity effect, where the $2\pi$ phase winding of the order parameter around the Fermi surface arises from the inherent wave function effect of the Dirac-like bands shown in Fig.~\ref{fig:figure1}(b). The authors further demonstrated that a quantum magnetic vortex can harbor a MZM \cite{Fu:2008aa}. Motivated by this idea, Bi$_2$Se$_3$ thin films have been successfully grown on a conventional $s$-wave superconductor NbSe$_2$, and the coexistence of topological surface states and superconductivity has been observed on the surface of this heterostructure \cite{Wang:2012ab}. Figs.~\ref{fig:figure1}(c)-(d) show scanning tunneling microscopy/spectroscopy (STM/STS) of a single magnetic vortex on the surface of Bi$_2$Te$_3$/NbSe$_2$ \cite{Xu:2014aa,Xu:2015aa}, where the symmetric zero-bias peak around the vortex center is a signature of a MZM \cite{Xu:2015aa}. Spin-selective Andreev reflections are detected in the zero-bias peaks using spin-polarized STM \cite{Sun:2016aa}, consistent with the theoretical prediction of the tunneling spectrum around a MZM \cite{He:2014aa}. 

Similar ideas have been extended to the heterostructure comprised of a semiconductor with spin-orbit coupling and a conventional superconductor \cite{Sau:2010aa,Lutchyn:2010aa,Oreg:2010aa}. Here the Zeeman field is essential in order to break time-reversal symmetry, leaving a single Fermi surface within the energy window of pairing potential \cite{Alicea:2012aa,Sato:2017aa}. The Zeeman field can be introduced into the heterostructure via magnetic proximity effect \cite{Sau:2010aa}, magnetic dopants \cite{Liu:2009aa,Yu:2010aa,Zhu:2011aa,Qin:2014aa}, or an external magnetic field \cite{Alicea:2010aa,Lutchyn:2010aa,Lutchyn:2011aa}. The last approach invokes a large Land\'e $g$ factor of the semiconductor \cite{Alicea:2010aa}. Because 2D superconductivity is easily destroyed by the orbit effect caused by perpendicular magnetic fields, 1D semiconductor/superconductor heterostructure that mimics the Kitaev chain attracts more attentions than its 2D counterpart \cite{Alicea:2011aa,Lutchyn:2018aa,Prada:2020aa}. Signatures of TSC and Majorana fermion have been unveiled in heterostructures consisting of conventional superconductors proximity coupled to 1D spin-orbit coupled semiconductor nanowires \cite{Mourik:2012aa,Rokhinson:2012aa,Das:2012aa,Albrecht:2016aa,Deng:2016aa,Nichele:2017aa,Lutchyn:2018aa}, magnetic atomic chains or islands  \cite{Nadj-Perge:2014aa,Ruby:2015aa,Menard:2017aa,Palacio-Morales:2019aa,Kezilebieke:2020aa}. However, experimental demonstration of Majorana fermion is far beyond definitive in view of the recent controversies surrounding the existence of MZMs in nanowire/superconductor heterostructures \cite{Zhang:2018aa}, and chiral Majorana edge modes in quantum anomalous Hall insulator-superconductor devices \cite{He:2017aa,Kayyalha:2020aa}. 

\subsection{Reciprocal-space proximity effect-induced TSC}
\label{subsec:reciprocal_proximity}
\begin{figure}[H]
\centering
\includegraphics[scale=1]{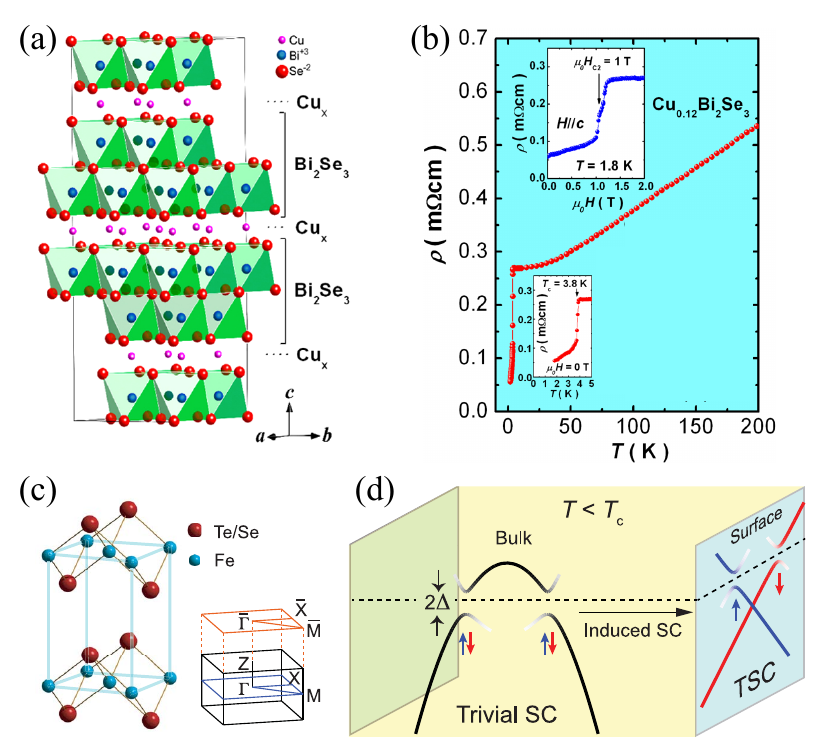}
\caption{(Color online) (a) Crystal structure of Cu$_x$Bi$_2$Se$_3$ realized by intercalating Cu atoms between quintuple layers of Bi$_2$Se$_3$. (b) Resistivity measured for $x = 0.12$, where superconducting transition occurs at $T_c= 3.8$ K. An upper critical field of $\mu_oH_{c2}$ $\sim 1$ T is observed. (c) Crystal structure and Brillouin zone (BZ) of Fe(Te, Se). (d) Schematic of the proximity effect between superconducting bulk states and topological surface states. (a) and (b) are reprinted from ref.~\cite{Hor:2010aa} with permission from American Physical Society. (c) and (d) are reprinted from ref.~\cite{Zhang:2018ab} with permission from AAAS.} 
\label{fig:figure2}
\end{figure}

By integrating the essential ingredients of a topological insulator/superconductor heterostructure, including Dirac-type surface sates and superconductivity in a single material, this constitutes the scheme of reciprocal-space or band proximity effect-induced TSC. Such an idea has been widely explored in doping carriers into topological insulators \cite{Hor:2010aa,Sasaki:2011aa,Levy:2013aa,Liu:2015aa}. For example, as shown in Figs.~\ref{fig:figure2}(a)-(b), Cu$_x$Bi$_2$Se$_3$ exhibits superconductivity  below $T_c \sim 3.8$ K for $0.12\le x \le 0.15$ \cite{Hor:2010aa}. In this mild doping regime, ARPES measurements show that the topological surface states survive up to the Fermi level \cite{Wray:2010aa}. By invoking the proximity effect between superconducting bulk states and topological surface states, this material serves as a promising platform for exploiting topological superconductivity and Majorana fermion \cite{Sasaki:2011aa,Hsieh:2012aa,Bay:2012aa,Levy:2013aa,Kirshenbaum:2013aa,Asaba:2017aa}. Although the origin of the ZBCP observed in point-contact measurements is still under active debates \cite{Sasaki:2011aa,Kirzhner:2012aa,Levy:2013aa}, the nuclear magnetic resonance (NMR) measurements of Knight shift \cite{Matano:2016aa} and specific heat measurements \cite{Yonezawa:2017aa} show that the spin-rotational symmetry is spontaneously broken below $T_c$ in Cu$_x$Bi$_2$Se$_3$, suggesting spin-triplet pairing and topological superconductivity.

Similar band proximity effect has been studied in iron-based superconductors \cite{Xu:2016aa,Shi:2017aa,Zhang:2017aa,Zhang:2018ab,Liu:2018aa,Hao:2018aa,Zhang:2019aa,Kong:2019aa,Wang:2020aa}. A typical example is FeTe$_{1-x}$Se$_x$, whose crystal structure is shown in Fig.~\ref{fig:figure2}(c). This material possesses topologically nontrivial band structures that are characterized by Dirac-type surface states for $x \sim 0.5$ \cite{Wang:2015aa,Wu:2016aa,Xu:2016aa}. Spin-resolved ARPES measurements in FeTe$_{0.55}$Se$_{0.45}$ reveal an isotropic superconducting gap with a size of $\Delta \sim 1.8$ meV in the spin-momentum-locking surface states caused by band proximity effect \cite{Zhang:2018ab}, as schematically depicted in Fig.~\ref{fig:figure2}(d), strongly indicating topological superconductivity. Moreover, a robust ZBCP is detected inside a surface magnetic vortex core in the STS \cite{Wang:2018aa,Machida:2019aa}, and the magnitude of the tunneling conductance is close to the conductance quantum of $2e^2/h$ \cite{Zhu:2020aa}. A more adequate review on realizing TSC and Majorana fermion in iron-based superconductors can be found in a separate review within this volume \cite{Ding:2022aa}. In view of its relative high critical temperature, iron-based superconductor serves as a superior condensed-matter platform for realizing and manipulating Majorana fermions. 

\section{Candidate systems for intrinsic topological superconductivity}
\label{sec:intrinsic}
Superconductivity and nontrivial topology of an intrinsic TSC emerge from the same electronic states. Intrinsic TSC is usually associated with unconventional pairing arising from strong electronic correlations. As illustrated in Eq.~(\ref{eq:gapequation}), if the many-electron correlation-dressed interaction matrix $V(\bm{k},\bm{k}')$ has an odd-parity structure with respect to $\bm{k}$ or $\bm{k}'$, the resulting superconducting state prefers spin-triplet pairing and $\Delta(\bm{k})$ possesses $2\pi$ phase winding around the Fermi surface. Because there are limited candidates of the intrinsic TSC, in this section, we review several representative examples ranging from 3D to 2D materials.

\subsection{Sr$_2$RuO$_4$}
Superconductivity was discovered in Sr$_2$RuO$_4$ in 1994 by Maeno \textit{et al} \cite{Maeno:1994aa}. Although this material shares similar tetragonal crystal structure with high transition temperature (high-$T_c$) Cu-based superconductors, Sr$_2$RuO$_4$ exhibits a much lower critical temperature of $T_c \sim 1.5$ K \cite{Mackenzie:1998aa,Mackenzie:1998ab}. The normal state of Sr$_2$RuO$_4$ does not show long-range magnetic orders. Nevertheless, strong ferromagnetic fluctuations are observed in low temperatures, likely preferring spin-triplet superconductivity \cite{Rice:1995aa,Mazin:1997aa}. Signatures of odd-parity spin-triplet  pairing in Sr$_2$RuO$_4$ are revealed in various types of experimental measurements including: (i) The NMR measurements show that the Knight shift remains constant \cite{Ishida:1998aa} or increase slightly \cite{Ishida:2015aa} (similar to the A phase of $^3$He \cite{Leggett:1975aa}) across $T_c$, suggesting that the spin susceptibility survives in the superconducting state, consistent with spin-triplet pairing; (ii) Spin relaxation is detected in muon spin rotation ($\mu$SR) measurements below $T_c$ \cite{Luke:1998aa}, indicating the existence of internal magnetic field, which breaks time-reversal symmetry of the superconducting state; (iii) Nonzero magneto-optic Kerr rotation in the superconducting state of Sr$_2$RuO$_4$ is another evidence of time-reversal symmetry breaking \cite{Xia:2006aa}; (iv) Phase-sensitive measurements reveal that the zero-field critical current of a $90$-degree corner Josephson junction lies  between the maximum and minimum critical currents achieved at finite magnetic fields \cite{Nelson:2004aa}, consistent with $p$-wave pairing; (v) Signatures of half-quantized vortices are observed in Sr$_2$RuO$_4$ in the presence of an oblique magnetic field \cite{Jang:2011aa}. These experimental observations single out the spin-triplet chiral $p$-wave pairing as the most likely pairing symmetry in Sr$_2$RuO$_4$ superconductor. Particularly, the recent quasi-particle tunneling spectroscopy measurements found signatures of edge states in Sr$_2$RuO$_4$ below $T_c$  \cite{Kashiwaya:2011aa}, which are further shown to be consistent with chiral $p$-wave superconductivity \cite{Kashiwaya:2011aa}.  Although these earlier experimental observations suggested that Sr$_2$RuO$_4$ is a chiral $p$-wave spin-triplet  superconductor, the recently observed deduction of Knight shift in the superconducting states of strained and untrained Sr$_2$RuO$_4$  \cite{Pustogow:2019aa} casts serious doubt on the previously believed odd-parity spin-triplet pairing. In fact, the precise pairing symmetry of Sr$_2$RuO$_4$ is still highly debated. Recent thermodynamic \cite{Benhabib:2021aa}, ultrasound \cite{Ghosh:2021aa}, and muon spin relaxation \cite{Grinenko:2021aa} measurements showed signatures of two-component order parameter in Sr$_2$RuO$_4$. Based on these experimentally revealed constraints, various theoretical proposals of even-parity spin-singlet pairing \cite{Gingras:2019aa,Romer:2019aa,Roising:2019aa,Suh:2020aa,Kivelson:2020aa} have also been proposed to understand the exotic superconducting properties of this material.

\subsection{Heavy-fermion material UTe$_2$}
Because spin-triplet pairing is commonly mediated by ferromagnetic fluctuation, heavy-fermion systems that contain $f$ electrons and possess both strong electronic correlations and magnetism are promising candidates for odd-parity spin-triplet superconductors. A typical heavy-fermion family that exhibits superconductivity is uranium-based compounds, such as UPt$_3$ \cite{Stewart:1984aa}, UGe$_2$ \cite{Saxena:2000aa}, URhGe \cite{Aoki:2001aa} and UCoGe \cite{Huy:2007aa}. There are various experimental signatures showing the coexistence of superconductivity and ferromagnetic fluctuation in these materials \cite{Aoki:2019aa}. Recently, superconducting transition temperature of $T_c\sim 1.6$ K has been observed in UTe$_2$ \cite{Ran:2019aa}, a new family member of uranium-based heavy-fermion compounds. In the phase diagram of uranium-based heavy-fermion superconductors, higher $T_c$ observed in UTe$_2$ than other family members is likely related to the fact that UTe$_2$ locates more closer to the critical ferromagnetic fluctuation point \cite{Ran:2019aa}.

Several types of experimental measurements show signatures of intrinsic topological superconductivity in UTe$_2$: (i) Temperature-independent NMR Knight shift across $T_c$ \cite{Ran:2019aa}; (ii) Extremely large anisotropic upper critical magnetic fields that are well in excess of the Pauli limit \cite{Aoki:2019ab}; (iii) The low-temperature magnetic behaviors manifest as a quantum critical ferromagnet with strong magnetic fluctuations \cite{Ran:2019aa}; (iv) A re-entrant superconducting phase at magnetic fields beyond 65 T \cite{Ran:2019ab}; (v) STS signatures of chiral in-gap states at step edges, resembling the Majorana boundary states in a TSC \cite{Jiao:2020aa}. All these experimental observations in together suggest UTe$_2$ is a promising intrinsic spin-triplet chiral $p$-wave TSC.

\subsection{Monolayer Pb$_{3}$Bi alloys}
\label{subsec:PbBiRG}
As discussed in the beginning of Sec.~\ref{sec:proximity}, nontrivial phase winding of superconducting order parameter can arise from either geometric phase of the single-particle wave function effect or many-electron correlation effect. For systems containing both effects, it is therefore essential to explore the interplay between them, and how they mutually influence each other in realizing intrinsic TSC. In 2019, Qin \textit{et al} proposed that the monolayer of Pb$_3$Bi grown on Ge(111) can harbor unconventional chiral $p$-wave superconductivity by taking advantage of the interplay between electron correlation and geometric phase \cite{Qin:2019aa}.
Before the theoretical proposal, superconductivity has been observed in ultra thin Pb film consisting of two atomic layers \cite{Qin:2009aa}, and later in one atomic Pb layer \cite{Zhang:2010aa}. Comparing to 3D bulk material, electronic screening in these 2D layers is dramatically reduced due to dimensional reduction, leading to an enhancement of electronic correlation. Moreover, the strong spin-orbit coupling of monolayer Pb$_3$Bi and substrate-induced inversion symmetry breaking result in non-vanishing geometric phase of the normal electronic states \cite{Qin:2019aa}.

\begin{figure}[H]
\centering
\includegraphics[scale=0.44]{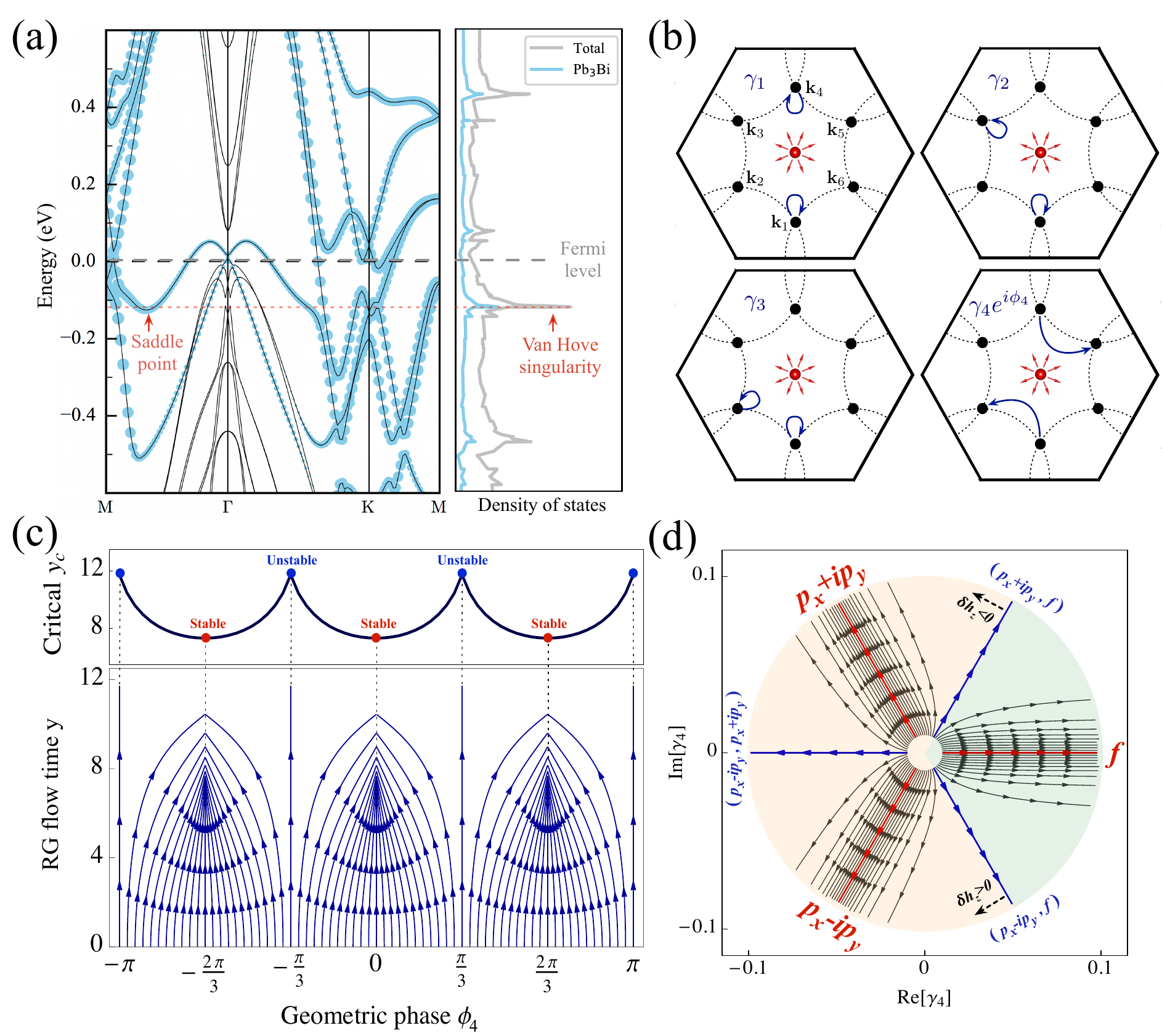}
\caption{(Color online) (a) Band structure and density of states of Pb$_3$Bi/Ge(111). Saddle point of the Rashba-split band is marked by red arrow and a type-II VHS emerges at $\sim$ 0.1 eV below the Fermi level. (b) Four distinct electron-electron scattering channels. A single monopole is schematically depicted at the BZ center, giving phase $\phi_4$ to $\gamma_{4}$. (c) RG flow of $\phi_4$ with stable fixed points highlighted by red dots. (d) Superconducting phase diagram spanned by $\gamma_4$. Red lines hight three stable fixed points that favor chiral $p_x \pm ip_y$ and $f$-wave pairing states.  (a)-(d) are reprinted from ref. \cite{Qin:2019aa} with permission from Springer Nature.} 
\label{fig:figure3}
\end{figure}

The band structure of Pb$_3$Bi/Ge(111) is shown in Fig.~\ref{fig:figure3}(a), where an extremely large Rashba-type spin-orbit coupling splitting and a saddle-like band structure emerges around the Fermi level. The density of states exhibits a type-II van Hove singularity (VHS) \cite{Yao:2015aa} emerging from the Rashba-split band characterized by non-vanishing geometric phase, which further contributes phase term to the electron-electron interaction \cite{Shi:2019aa}. 
Within saddle-point patch approximation, an effective interacting model is developed. As depicted in Fig.~\ref{fig:figure3}(b), $\gamma_4$ is dressed by a phase factor of $\phi_4$ originating from geometric phase of the wavefunction. A couple of complex renormalization group (RG) flow equations are derived as \cite{Qin:2019aa}
\begin{equation}
\begin{aligned}
\frac{ \text{d} \gamma_1}{ \text{d}y}&=2(\eta-y)\gamma_1^2 -8\gamma_2\gamma_3-4y |\gamma_4|^2,\\
\frac{ \text{d} \gamma_2}{ \text{d}y}&= (\eta'_1-\eta_1-2)\gamma_2^2-4\gamma_1\gamma_3-2 \gamma_3^2+\eta'_1 |\gamma_4|^2,\\
\frac{ \text{d} \gamma_3}{ \text{d}y}&= -4(\gamma_1\gamma_2+\gamma_2\gamma_3)+(\eta'_1-\eta_1) \gamma_3^2+\eta'_1 |\gamma_4|^2, \\
\frac{ \text{d} \gamma_4}{ \text{d}y}&=-4y \gamma_1\gamma_4+2y (\gamma_4^*)^2+2 \eta'_1(\gamma_2+\gamma_3)\gamma_4,
\end{aligned}
\label{eq:equation3}
\end{equation}
where the relevant coefficients can be found in ref.~\cite{Qin:2019aa}. By solving these differential equations, pairing instability occurs at a critical temperature $T_c \sim 4\Lambda e^{-y_c}$, where $\Lambda$ is the ultraviolet energy cutoff and $y_c$ is the critical RG flow time.

As shown in Figs.~\ref{fig:figure3}(c)-(d), the renormalized geometric phase flows to three stable fixed points at $\pm 2\pi/3$ and $0$, preferring ($p_x\pm ip_y$)-wave and $f$-wave pairings, respectively. Given the robustness of the stable points, two thirds of the phase diagram shown in Fig.~\ref{fig:figure3}(d) are able to harbor chiral $p$-wave superconductivity. The pairing mechanism can arise either from  electron-phonon interaction or from electron-electron repulsive interaction. Moreover, the critical RG flow time $y_c$ shown in Fig.~\ref{fig:figure3}(c) exhibits minima at stable fixed points, suggesting enhanced superconducting critical temperature. Overall, monolayer Pb$_3$Bi serves as a promising material for realizing 2D intrinsic chiral $p$-wave TSC.

\subsection{Graphene-based systems} 
Graphene is a monolayer isolated graphite consisting of carbon atoms arranged on a honeycomb lattice. The low-energy band structure of graphene possesses a linear dispersion associated with Dirac-like electronic excitations \cite{Castro-Neto:2009aa}. By doping graphene to the VHS, it was proposed that chiral $d$-wave superconductivity can emerge from repulsive electron-electron interaction \cite{Nandkishore:2012aa}. In experiments, it is hard to reach the VHS of graphene through conventional electric gating. Superconductivity is observed in heavily doped monolayer graphene with alkali adatoms, where both of the conventional electron-phonon interaction and pure electron-electron interaction are invoked to explain the observed superconductivity \cite{McChesney:2010aa,Profeta:2012aa,Ludbrook:2015aa}. 

By stacking two sheets of graphene with a relative $\sim$$1.1^{\circ}$ twisting angle (magic angle), Bistritzer and MacDonald found the corresponding moir\'e mini bands are flat near Fermi level \cite{Bistritzer:2011aa}, leading to VHS in the density of states. The quenched kinetic energy in magic-angle twisted bilayer graphene (MATBG) strongly enhances the electronic correlation. In 2018, Cao \textit{et al} discovered correlated insulating state at half filling in MATBG \cite{Cao:2018ab} and superconductivity upon slightly doping away from half filling \cite{Cao:2018aa}. The phase diagram of MATBG resembles that of Cu-based high-$T_c$ superconductor, where superconductivity emerges upon doping away from Mott insulator \cite{Dagotto:1994aa}. In contrast to cuprates, MATBG has an exceptional property of low carrier density that is tunable via external electrostatic gating, making it an ideal platform for systematically studying strong correlation effects  \cite{Yankowitz:2019aa,Lu:2019aa,Balents:2020aa,Stepanov:2020aa,Saito:2020aa,Wong:2020aa,Zondiner:2020aa,Liu:2021aa,Stepanov:2021aa,Saito:2021aa,Choi:2021aa,Choi:2021ab,Jaoui:2022aa}. To date, the underlying mechanism of superconductivity in MATBG remains under active debates; both the conventional electron-phonon interaction mediated $s$-wave pairing \cite{Wu:2018aa,Choi:2018aa,Lian:2019aa,Angeli:2019aa,Lewandowski:2021aa} and strong electron-electron interaction drived unconventional pairing \cite{Guo:2018aa,Dodaro:2018aa,Liu:2018ab,Isobe:2018aa,Kennes:2018aa,Sherkunov:2018aa,Xie:2019aa,Kerelsky:2019aa,Choi:2019aa,Jiang:2019aa,Gonzalez:2019aa,Roy:2019aa,Huang:2019aa,Ray:2019aa,You:2019aa,Khalaf:2020aa,Qin:2021aa,Cea:2021aa,Huang:2021aa} have been theoretically proposed. In particular, there are theoretical studies showing that superconductivity in MATBG is topologically nontrivial \cite{Xu:2018aa,Fidrysiak:2018aa,Chew:2021aa}.

Mirror symmetric magic-angle twisted trilayer graphene (MATTG) consists of three layers of graphene stacked in a mirror symmetric configuration with twisted top and bottom layers \cite{Park:2021aa}. MATTG possesses nearly identical flat bands to MATBG at a magic angle that differs from $1.1^{\circ}$ by a factor of $\sqrt{2}$  \cite{Khalaf:2019aa,Mora:2019aa,Carr:2020aa,Zhu:2020ab,Lei:2020aa}. The two systems share similar phase diagrams upon varying carrier density, including the patterns of broken flavor symmetries and the emergence of superconductivity away from the correlated insulating states \cite{Park:2021aa,Hao:2021aa}. 
A remarkable difference between MATBG and MATTG superconductors is their responses to in-plane magnetic fields. The critical in-plane magnetic field of MATBG is compatible with Pauli paramagnetic limit \cite{Cao:2021ab}, whereas superconductivity in MATTG survives to in-plane magnetic fields that are well in excess of this limit \cite{Cao:2021aa}. Such distinct in-plane critical magnetic fields in MATBG and MATTG can be well explained by invoking spin-triplet pairing in both systems \cite{Qin:2021ab}, where the orbit effect that breaks Cooper pairing in MATBG is quenched by the inherent mirror symmetry in MATTG \cite{Qin:2021ab,Lake:2021aa}. These studies suggest twisted graphene multilayers are promising candidates for harboring spin-triplet superconductivity. 

Superconductivity was recently observed in electric-gated untwisted rhombohedral trilayer graphene (RTG) \cite{Zhou:2021aa} and Bernal-stacking bilayer graphene \cite{Zhou:2022aa}. RTG exhibits two superconducting domes upon varying carrier density. The higher $T_c$ dome emerges from a paramagnetic normal state with annular Fermi sea \cite{Zhou:2021aa,Zhou:2021ab}, which is demonstrated to be able to harbor unconventional chiral-$p$ or $d$-wave superconductivity \cite{Chatterjee:2021aa,Ghazaryan:2021aa,Cea:2022aa,Qin:2022aa}. The other superconducting dome emerges from the spin-polarized, valley-unpolarized half-metallic state, and survives to in-plane magnetic fields far beyond the Pauli limit, indicating spin-triplet pairing \cite{Zhou:2021aa}. For Bernal bilayer graphene, superconducting phase transition occurs only at finite in-plane magnetic fields \cite{Zhou:2022aa}, an essential character of spin-triplet superconductivity. Overall, the (near) flat bands in twisted and untwisted graphene multilayers strongly enhance their electronic correlations, providing the ground for developing unconventional superconductivity, serving as candidate systems for realizing 2D intrinsic TSC. 

\section{Coexistence of superconductivity and topological bands}
\label{sec:coexist}
In this section, we review several recently proposed and/or discovered 2D materials showing coexistence of superconductivity and topological bands. Based on the principle of reciprocal-space proximity effect discussed in Sec.~\ref{subsec:reciprocal_proximity}, these materials are candidates for hosting 2D or 1D topological superconducting states.

\subsection{Monolayer Pb$_{3}$Bi }
\begin{figure}[H]
\centering
\includegraphics[scale=0.47]{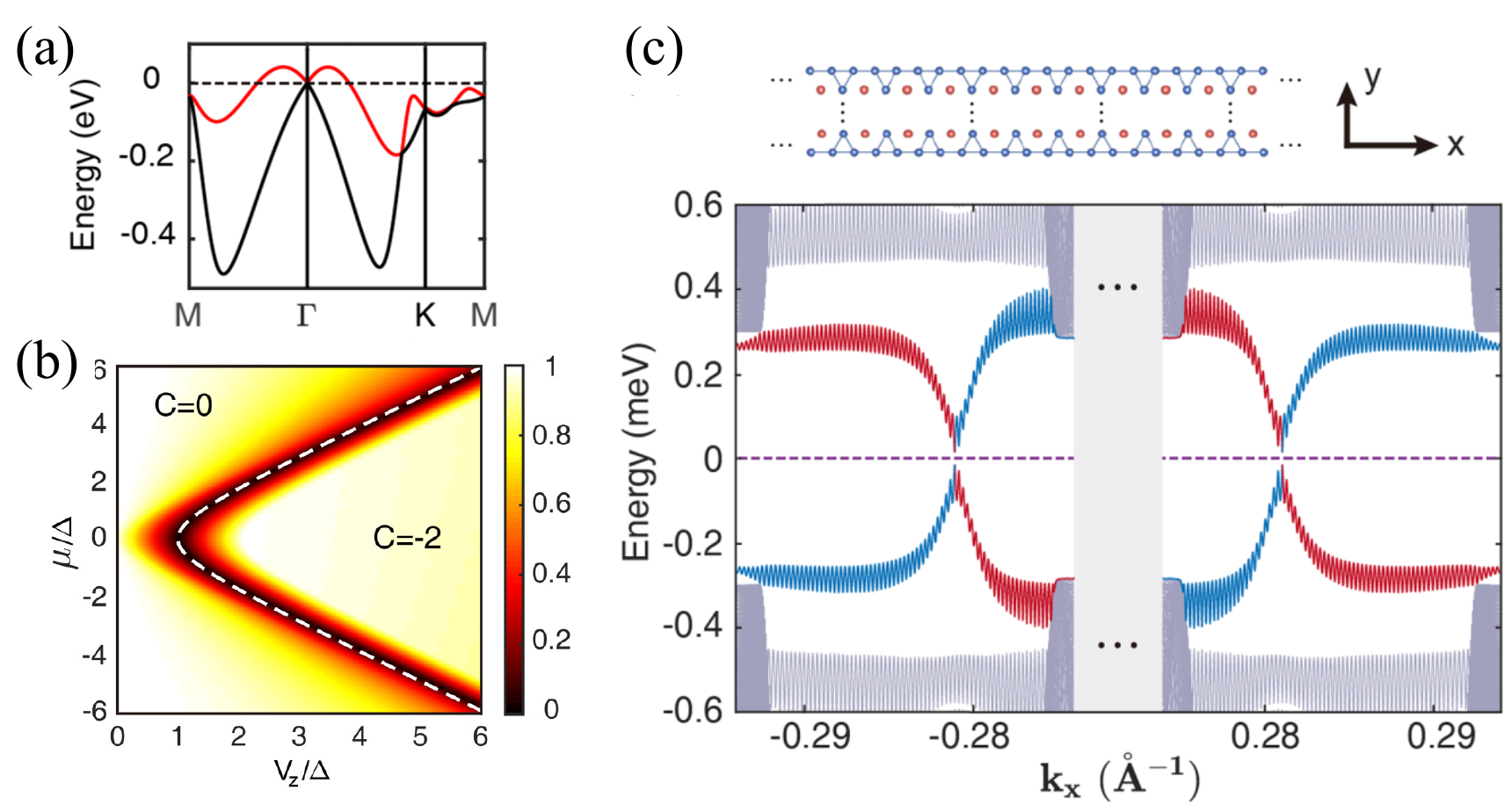}
\caption{(Color online) (a) Rashba bands calculated from the tight-binding model. (b) BdG quasiparticle band gap (color scale) versus Zeeman field $V_z$ and chemical potential $\mu$ with pairing potential $\Delta = 0.3$ meV. (c)  Quasiparticle bands of the ribbon structure depicted in the upper panel, where $\Delta = 0.3$ meV and $V_z = 1.5$ meV. Oscillation of chiral Majorana edge modes is caused by finite size effects \cite{Zhou:2008aa,Wada:2011aa}. (a)-(c) are reprinted from ref.~\cite{Sun:2021aa} with permission from American Physical Society.} 
\label{fig:figure4}
\end{figure}

In Sec.~\ref{subsec:PbBiRG}, monolayer Pb$_{3}$Bi grown on Ge(111) is shown to be an intrinsic 2D TSC at VHS band filling. By tuning the Fermi level to the Dirac point at the BZ corner, this material was also proposed to realize TSC via band proximity effect, where superconductivity and topology arise from different electronic bands \cite{Sun:2021aa}. An effective tight-binding model was developed to describe the electronic structure of Pb$_3$Bi/Ge(111) \cite{Sun:2021aa}. By comparing Fig.~\ref{fig:figure3}(a) and Fig.~\ref{fig:figure4}(a), the Rashba bands obtained from density functional theory calculation are well reproduced by the model calculation, including the giant Rashba-type splitting, type-II VHS, and the Dirac-like bands around the BZ corner. By invoking the conventional electron-phonon interaction-mediated spin-singlet pairing, as illustrated in Fig.~\ref{fig:figure4}(b), a phase transition from topologically trivial superconducting state with Chern number $C = 0$ to topological superconducting state with $C = -2$ occurs when the Bogoliubov–de Gennes (BdG) quasiparticle band gap is closed. It was further shown that the $C = -2$ state is protected by mirror symmetry \cite{Sun:2021aa}. The lower panel of Fig.~\ref{fig:figure4}(c) shows the BdG quasiparticle bands of the ribbon structure depicted in the upper panel of Fig.~\ref{fig:figure4}(c). There are two chiral Majorana edge modes propagating along the same direction at each edge, consistent with the Chern number calculation.

Experimental realization of TSC and chiral Majorana fermions in monolayer Pb$_{3}$Bi is highly promising due to the following reasons: (i) Both the VHS and K-point Dirac cone are close to Fermi level and accessible via chemical doping or ionic liquid gating \cite{Ye:2010aa}; (ii) Pairing potential with a magnitude $\Delta = 0.3$ meV is easily obtained because the experimentally observed superconducting gaps for Pb thin films and related systems range from 0.3 meV to 1 meV \cite{Qin:2009aa,Zhang:2010aa,Sekihara:2013aa,Yamada:2013aa,Nam:2016aa}; (iii) The large Land\'e $g$-factor makes it possible to convert the system into a TSC by applying a small external magnetic field that does not completely suppress superconductivity. For example, by taking $g\sim100$ \cite{Lei:2018aa}, the topological phase transition occurs at $V_z \sim \Delta$, amounting to $B_{\bot} = 0.1 $ T. This value is smaller than the critical magnetic field $B_{c\bot}$ of Pb thin films, ranging from 0.15 T to 1 T \cite{Zhang:2010aa,Yamada:2013aa}. Overall, monolayer Pb$_3$Bi alloy is an appealing platform for realizing 2D TSC.

In addition to the superconducting state, the structural, electronic, and topological properties of Pb$_3$Bi/Ge(111) have been systematically studied using first-principles calculations \cite{Li:2020aa}. Three different structural phases labeled respectively by T$_1$, H$_3$ and T$_4$ were shown to be energetically nearly degenerate. Moreover, the electronic structures of H$_3$ and T$_4$ phases are topologically nontrivial and can harbor quantum spin Hall effect \cite{Li:2020aa}. 

\subsection{2D CoX (X = As, Sb, Bi) }
Discoveries of high-$T_c$ superconductivity in Cu- and Fe-based superconductors have generated tremendous interests over the last decades \cite{Damascelli:2003aa,Paglione:2010aa,Si:2016aa}. Among the Fe-based superconductors, bulk FeSe possesses relatively simple crystal structure and exhibits a critical temperature $T_c \sim 8 $ K at ambient pressure \cite{Hsu:2008aa}, which can be enhanced to 37 K under high pressure \cite{Medvedev:2009aa}. Importantly, it provides an elemental building block, namely, the FeSe monolayer, which can be placed on proper substrates \cite{Wang:2012aa,Peng:2014aa,Ding:2016aa} or stacked into superlattice structures \cite{Guo:2010aa,Lu:2015ab} and then results in substantially enhanced $T_c$’s up to tens of Kelvin. Extensive studies have been carried out to search for potential high-$T_c$ superconductors beyond Cu- and Fe-based families. A comprehensive project screened over 1000 candidate materials, but no high-$T_c$ superconducting material was indentified beyond Cu- and Fe-based families \cite{Hosono:2015aa}. In those earlier searches, only systems with intrinsically layered structures were considered. More recent studies have revealed that 2D materials with no layered bulk phase can also be stabilized in the monolayer or few-layer regime \cite{Zhu:2017aa,Lucking:2018aa}, effectively broadening the space of 2D materials as candidate high-$T_c$ superconductors.

\begin{figure}[H]
\centering
\includegraphics[scale=0.5]{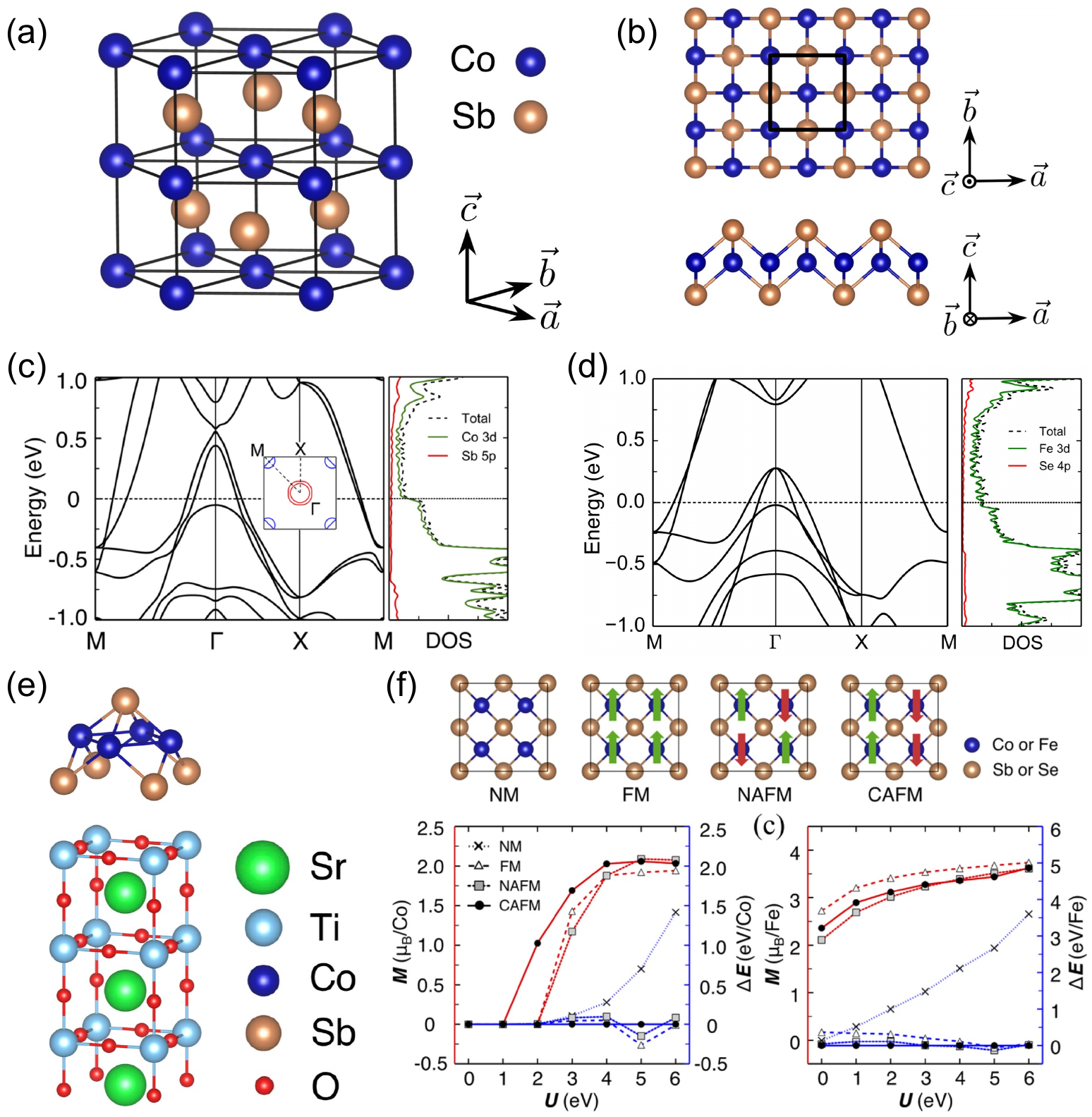}
\caption{(Color online) (a) and (b) Schematic atomic structures of NiAs-type bulk CoSb and PbO-type monolayered CoSb. (c) and (d) Electronic structures (left panel) and density of states (right panel) of freestanding CoSb and FeSe monolayers. (e) Most stable structure of CoSb monolayer on the SrTiO$_3$(001) substrate, where the Sb atoms sit right above the Ti atoms. (f) Four commonly considered magnetic configurations of CoSb/STO or FeSe/STO (top panel). Magnetic moment $M$ (red) and relative energy $\Delta E$ (blue) of the four magnetic configurations versus the on-site Hubbard $U$ for the two systems. This figure is reprint from ref.~\cite{Ding:2020aa} with permission from American Physical Society.} 
\label{fig:figure5}
\end{figure}

In 2020, Ding \textit{et al} \cite{Ding:2020aa} identified the monolayered CoSb to be an attractive candidate for harboring high-$T_c$ superconductivity. Their prediction is initially guided by the isovalency rule, namely, keeping the same number of valence electrons as FeSe but tuning other physical factors. Although the bulk structure of CoSb is hexagonal and non-layered, as shown in Fig.~\ref{fig:figure5}(a), its freestanding monolayer can be stabilized to the tetragonal structure depicted in Fig.~\ref{fig:figure5}(b). Particularly, monolayer CoSb and FeSe share an identical crystal structure and similar electronic bands, as illustrated in Figs.~\ref{fig:figure5}(c) and (d). For monolayer FeSe-based superconductors, the dominant pairing mechanism is still under active debate, with antiferromagnetic spin fluctuations \cite{Scalapino:2012aa,Chubukov:2012aa}, electron-phonon coupling (EPC) \cite{Lee:2014aa,Gerber:2017aa,Zhang:2019ab}, and the cooperative effect of the two to be frequently invoked \cite{Lee:2014aa,Lee:2018aa}. By employing the EPC strength as an indicator for superconductivity, Ding \textit{et al} \cite{Ding:2020aa} found that the freestanding CoSb monolayer ($T_c = 0.9$ K) possesses higher critical temperature than freestanding FeSe monolayer ($T_c = 0.5$ K). In analogy to FeSe, monolayer CoSb can be supported on SrTiO$_3$(001) [CoSb/STO, Fig.~\ref{fig:figure5}(e)], offering promising alternative platforms for realizing high-$T_c$ superconductivity under the regulation of substrates including significant charge transfer \cite{He:2013aa,Tan:2013aa} and the longitudinal optical phonons of the STO penetrating into the overlayers \cite{Zhang:2016aa}. In contrast to FeSe/STO, CoSb/STO shows a much weaker tendency toward developing magnetization, and exhibits completely different magnetic properties, as shown in Fig.~\ref{fig:figure5}(f). In addition, other cobalt pnictides, such as CoAs and CoBi monolayers, have also been demonstrated as candidate 2D high-$T_c$ superconductors, even though their bulk phases have no resemblance to layering either \cite{Gao:2021aa}. These findings offer new attractive candidate systems beyond the well-known highly crystalline 2D superconductors \cite{Saito:2016ab}. 

Motivated by the theoretical prediction \cite{Ding:2020aa}, several experimental groups have made efforts to study this system, with preliminary confirmative findings. Specifically, Xue and collaborators fabricated CoSb films in orthogonal structure on STO and observed symmetric gap around the Fermi level with coherence peaks at 7 meV as well a diamagnetic transition at 14 K, indicative of superconductivity \cite{Ding:2019aa}. CoSb$_{1-x}$ nanoribbons with quasi-one-dimensional stripes on STO were also fabricated, showing signatures of Tomonaga-Luttinger liquid state \cite{Lou:2021aa}. 

\begin{figure}[H]
\centering
\includegraphics[scale=0.57]{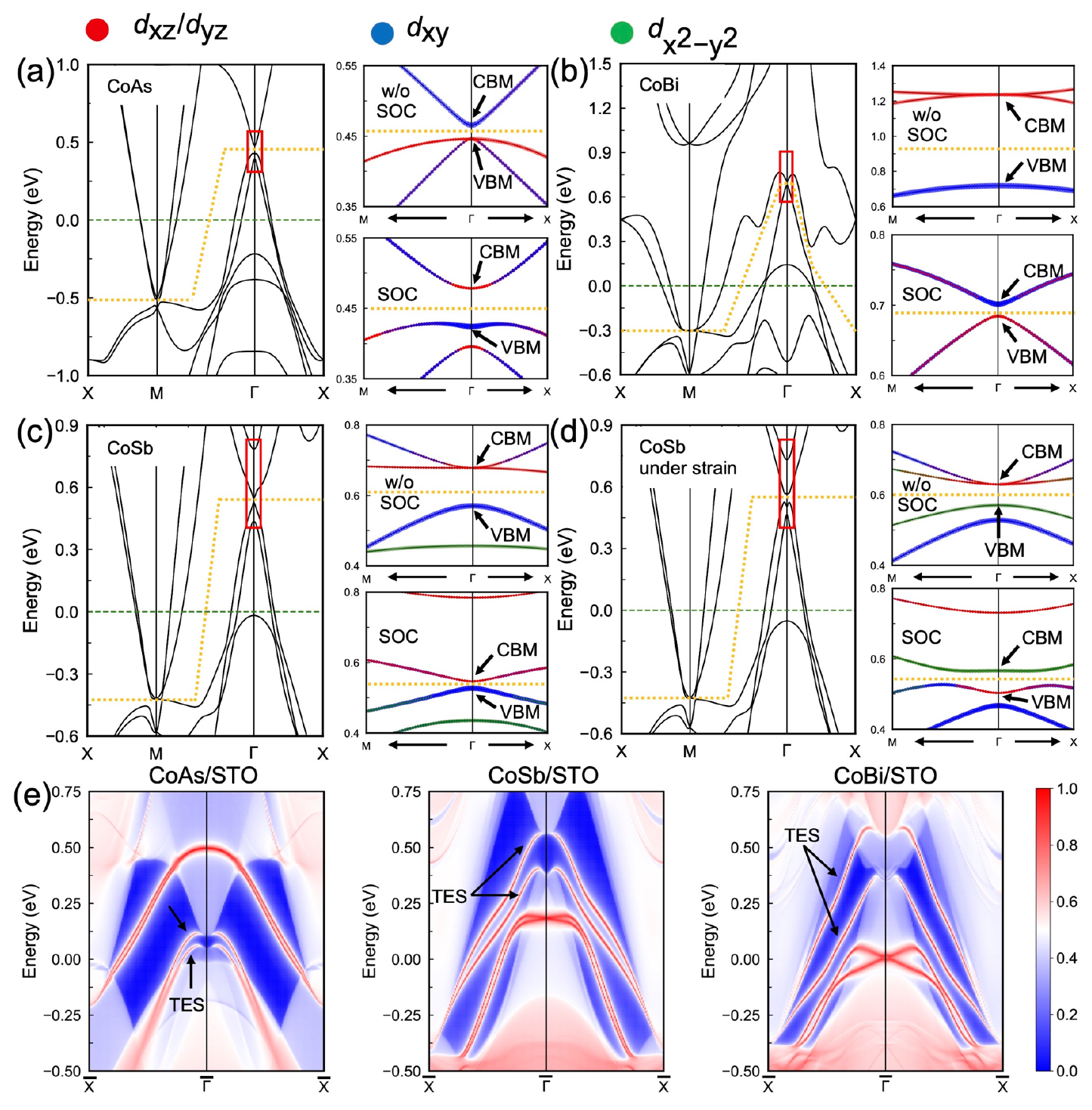}
\caption{(Color online) (a)-(d) The left panels are band structures of freestanding CoAs, CoBi, CoSb monolayers and CoSb monolayer with tensile biaxial strain of 0.7\%, respectively. These results are calculated with SOC. The right panels show the enlarged and orbital-resolved band structures within the regimes marked by solid red boxes in the left panels. The top and bottom panels plot the results calculated without and with SOC, respectively. The radii of the red, blue, and green dots indicate the spectral weights of different $d$ orbitals of Co atoms. The orange dashed lines in (a)-(d) correspond to the curved chemical potentials. (e) Topological edge states (TESs) of CoX/STO along the [100] edge. The warmer colors denote higher local density of states, and the blue regions denote the bulk band gaps. The Fermi levels are all set at zero. (a)-(e) are modified after ref.~\cite{Gao:2021aa} with permission from American Chemical Society.} 
\label{fig:figure6}
\end{figure}

Beyond superconductivity, the normal-state band topology of CoX monolayers has been systematically studied \cite{Gao:2021aa}. First-principles calculations of the band structures of freestanding CoX monolayers are shown in Figs.~\ref{fig:figure6}(a)-(c). In the absence of SOC, the local gaps around $\Gamma$ point between the conduction band minimum (CBM) and valence band maximum (VBM) are 20, 108, and 517 meV for CoAs, CoSb, and CoBi, respectively. In the presence of SOC, band inversion occurs at $\Gamma$ point for CoAs [Fig.~\ref{fig:figure6}(a)] and CoBi [Fig.~\ref{fig:figure6}(b)], indicating both systems are topologically nontrivial, as also confirmed by the calculated topological invariant $Z_2 = 1$. In contrast, the SOC in CoSb [Fig.~\ref{fig:figure6}(c)] is not strong enough to close and reopen the band gap. Nevertheless, a moderate tensile biaxial strain of 0.7\% can reduce the band gap to 59 meV, which can then be closed and reopened by the SOC [Fig.~\ref{fig:figure6}(d)], driving the system into a topologically nontrivial phase. The topological properties of CoX/STO have also been investigated. It turns out that each CoX monolayer possesses nontrivial band topology under the lattice constant of STO substrate and the influence of the STO bands on the topology of CoX is negligible. These results suggest that CoX/STO systems are topologically nontrivial, as characterized by the odd $Z_2$ invariants and robust edge states shown in Fig.~\ref{fig:figure6}(e). 

Since CoX/STO systems are able to harbor both high-$T_c$ superconductivity and nontrivial band topology, by further invoking the reciprocal-space proximity effect between 2D bulk superconducting sates and topological edge states, these systems are new promising candidates for realizing 1D topological superconductivity.

In addition to CoX systems, there are also other 2D candidate systems that are highly like to support the coexistence of superconductivity and nontrivial band topology. One typical example is PdTe$_2$ grown on STO substrate, where robust superconductivity was observed down to bilayer thickness of PdTe$_2$ \cite{Liu:2018ac}. Nevertheless, experimental confirmation of nontrivial band topology in PdTe$_2$/STO is challenging because the predicted topological edge states turn out to be heavily overlapped with the 2D bulk states \cite{Liu:2018ac}. Another example is monolayer W$_2$N$_3$, which can be mechanically exfoliated from its van der Waals bulk counterpart. Based on the fully anisotropic Migdal-Eliashberg formalism calculation, this material is unveiled to exhibit superconductivity below $T_c\sim21$ K associated with a zero-temperature superconducting gap of $\sim$ 5 meV \cite{You:2021aa,Campi:2021aa}. The topological helical edge states emerge at 0.5 eV above the Fermi level, where superconductivity was shown to be persistent \cite{You:2021aa,Campi:2021aa}. 

\subsection{Stanene atomic layers}
Stanene was proposed to be a large-gap quantum spin Hall insulator \cite{Xu:2013aa}, which has stimulated extensive efforts on synthesizing stanene films under diverse conditions and subsequently characterizing their emergent properties. Monolayered stanene was first successfully fabricated on the Bi$_2$Te$_3$(111) substrate by molecular beam epitaxy \cite{Zhu:2015aa}, and later also on other substrates, such as Sb(111) \cite{Gou:2017aa} and InSb(111) \cite{Xu:2018ab}. Such monolayered stanene films with compressed strain by the substrates were shown to exhibit only a trivial band structure. An advance in realizing nontrivial band topology was achieved by growing stanene on the Cu(111)-($2\times 2$) surface, where an inverted band order was observed in the ultraflat yet metastable stanene films \cite{Deng:2018aa}. Indications of edge states of monolayered stanene grown on the InSb(111)-($3\times 3$) substrate have been reported, even though the films contain pronounced defects \cite{Zheng:2019aa}. An unexpected high-buckled $\sqrt{3}\times \sqrt{3}$ stanene grown on the Bi(111) substrate has also shown a gap shape near the Fermi level and exhibited the existence of the edge states, which was further theoretically confirmed to be topologically nontrivial \cite{Song:2021aa}. At a separate front, superconductivity of few-layer stanene grown on PbTe(111) has been detected using transport measurements \cite{Liao:2018aa}, and more intriguingly, type-II Ising pairing (which will be described later) has been proposed to interpret the observed unusually large in-plane upper critical fields beyond the Pauli paramagnetic limit \cite{Wang:2019aa,Falson:2020aa}. These exciting developments demonstrate that stanene can serve as an ideal platform for studying the interplay between the nontrivial band topology and superconductivity, two central ingredients that can be further explored for realization of 2D topological superconductivity by invoking band proximity effect. 

\begin{figure}[H]
\centering
\includegraphics[scale=0.57]{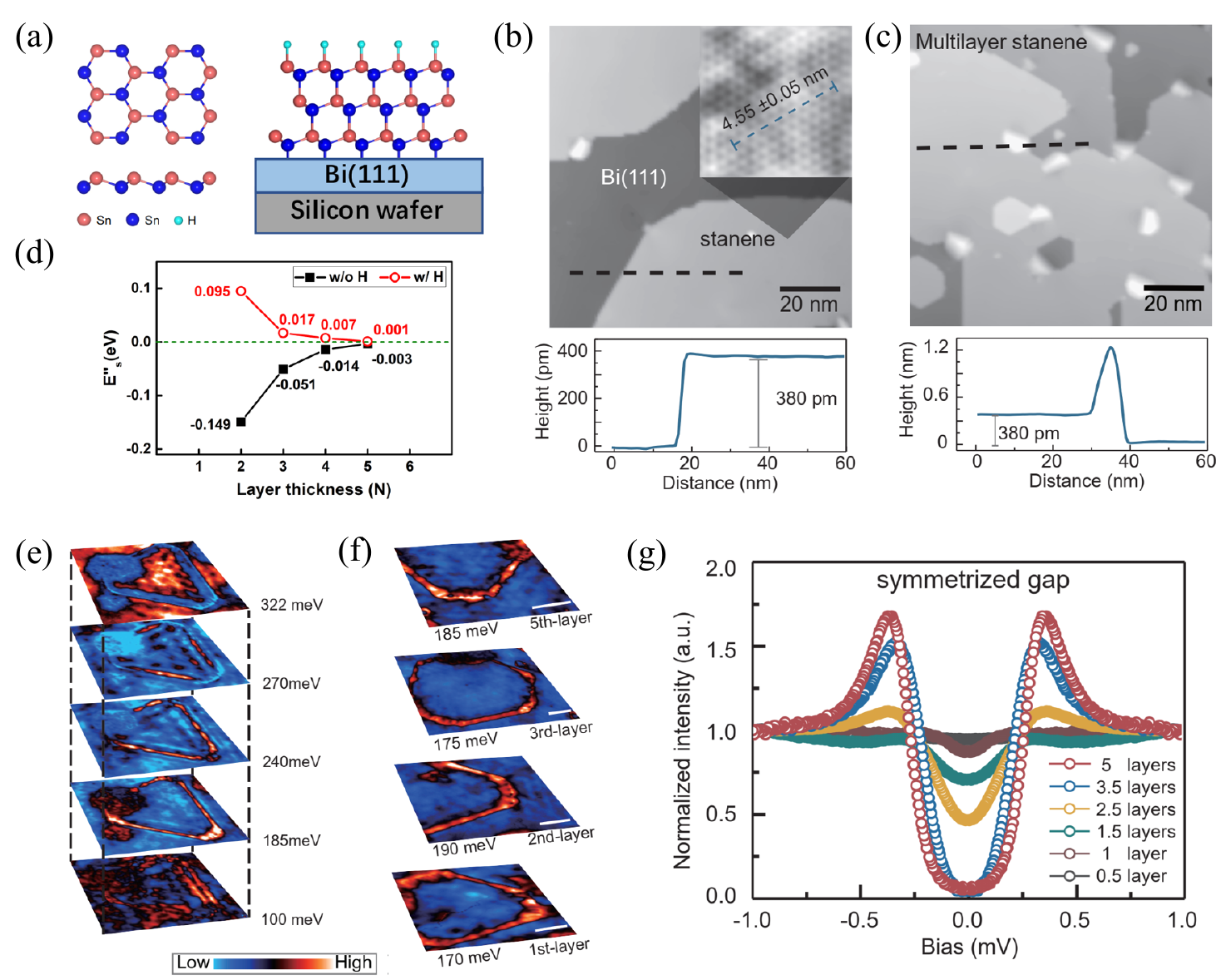}
\caption{(Color online) Coexistence of the robust edge states and superconductivity in one- to five-layer stanene films grown on Bi(111). (a) Schematics of a freestanding monolayer stanene (left) and a sample structure of four-layer stanene/Bi(111)/Si with the top surface of stanene passivated by hydrogen (right). (b) Topography of the first-layer stanene films on Bi(111), with a profile shown at the bottom depicting the height along the black dotted line. Inset: atomically resolved image taken on the stanene film. (c) Topography of multilayer stanene ($\sim$ 2.5 layers) on Bi(111), with the corresponding height profile shown below. (d) The second-order differences of the formation energies of N-layer stanene films as a function of the thickness of the stanene layers. (e) dI/dV mappings of a fourth-layer stanene island at different energies. (f) dI/dV mappings of different-layered stanene islands taken at the energy of the respective bulk dip minimum. (g) Layer-dependent superconducting gap of stanene films. (b)-(g) are from ref.~\cite{Zhao:2022aa} with permission from American Physical Society.} 
\label{fig:figure7}
\end{figure}

A major obstacle severely limiting the exploration of stanene is that the overall quality of such stanene films is still far from satisfactory, such as containing multi-domains \cite{Zhu:2015aa} or pronounced defects \cite{Zheng:2019aa}. A recent theoretical work reported that a stanene monolayer obeys different atomistic mechanisms when grown on different Bi$_2$Te$_3$(111)-based substrates, and in particular, the Bi(111)-bilayer precovered Bi$_2$Te$_3$(111) substrate was predicted to strongly favor the growth of single crystalline stanene \cite{Zhang:2018ac}. This prediction has been largely supported by the latest experimental demonstration of high-quality few-layer stanene grown on the Bi(111) substrate \cite{Zhao:2022aa}. As shown in Figs.~\ref{fig:figure7}(a)-(c), one- to five-layer stanene films with high quality were successfully fabricated on the Bi(111) films that were first grown on a silicon wafer, where the stanene films are stable at room temperature. The topmost surfaces of such stanene films are saturated by hydrogen atoms, similar as shown in previous studies \cite{Zhu:2015aa,Zang:2018aa}. During the growth of each layer of stanene, the most important processes are low-temperature deposition of Sn atoms and sufficient surface passivation of the films by the residual hydrogen. The systematic first-principles calculations further reveal that the surface passivation of the growth front is essential in achieving layer-by-layer growth of the high-quality stanene films, with the hydrogen functioning as a surfactant \cite{Copel:1989aa,Zhang:1997aa}. Specifically, as shown in Fig.~\ref{fig:figure7}(d), the second-order difference of the formation energy of an N-layer stanene film ($E^{''}_{s}(N)$) with hydrogen passivation is equal or larger than zero, indicating that the film is stable; in contrast, $E^{''}_{s}(N)$ without hydrogen passivation is negative, indicating that the film is unstable. 

With these high-quality stanene films, the exotic properties have been further experimentally observed, including the long-sought edge states and superconductivity. First, the \textit{in situ} STM/STS measurements identify the enhanced intensity of local density of states at two different zigzag edges of the stanene islands with different thicknesses, signifying the existence of the robust edge states, as shown in Figs.~\ref{fig:figure7}(e) and (f). First-principles calculations further show that all these films on Bi(111) have well-defined continuous gaps across the whole BZ with the inclusion of SOC, yielding an extraordinarily robust nontrivial topological invariant $Z_2$ that is impervious to the layer thickness (see Table \ref{tab1}). The physical origin of the robust nontrivial topology is the consequence of interfacial coupling with the Bi(111) substrate. Qualitatively, the Bi(111) substrate, with inherently strong SOC, is able to promote the nontrivial topology in the few-layer stanene via effective proximity effects. Furthermore, possible origins for topologically trivial edge states such as dangling bonds or H-passivation at the edges can be excluded by a comparative experiment of growing stanene on Bi$_2$Te$_3$. Next, clear superconducting gaps were detected on the stanene islands where the robust edge states were also observed. Fig.~\ref{fig:figure7}(g) shows the layer-dependent superconductivity of stanene films measured at 400 mK, exhibiting a wider and deeper superconducting gap with a thicker layer. In particular, the existence of a superconducting gap in monolayer stanene was observed for the first time, which might be enabled by the higher charge transfer level from the Bi(111) substrate. Overall, the coexistence of nontrivial topology and intrinsic superconductivity renders stanene a promising candidate for realizing 2D topological superconductivity in a simple single-element system. As aforementioned, such topological superconductivity originates from band proximity effect that has been experimentally demonstrated in several 3D systems \cite{Yin:2015aa,Zhang:2018ab,Wang:2018aa,Wang:2020aa,Zhang:2019aa,Liu:2018aa}. 

\begin{table}[H]
\footnotesize
\caption{$Z_2$ topological invariants of different-layered stanene films under different conditions. The table is adapted from ref.~\cite{Zhang:2018ac} with permission from American Physical Society.}
\label{tab1}
\tabcolsep 10pt 
\doublerulesep 0.025pt \tabcolsep 4pt 
\begin{tabular}{ l c c c c c}
\toprule
~~~~~Conditions & 1-layer & 2-layer & 3-layer & 4-layer & 5-layer \\
\hline
w/o Bi  ~~~~~~~~w/o H & 1 & 0 &1 &0 &1 \\
w/o Bi  ~~~~~~~~w/ H & 0 & 0 &1 &1 &1 \\
w/ Bi  ~~~~~~~~~~w/o H & 1 & 1 &1 &1 &1 \\
\begin{tabular}{@{} l @{} l}w/ Bi ~~~~~~~~~ w/ H  \\ (experimental condition) \end{tabular}
& 1 & 1 &1 &1 &1 \\
\bottomrule
\end{tabular}	
\end{table}

In order to realize topological superconductivity in these systems that harbor both superconductivity and topological bands via reciprocal-space or band proximity effect, one common crucial condition is the emergence of Dirac-cone-like electronic bands within the energy window of the superconducting gap. Similarly, this condition should also be fulfilled in other systems, such as monolayered Fe(Te, Se), gated WTe$_2$ monolayer, and the IrTe$_2$/In$_2$Se$_3$ heterobilayer, where the coexistence of superconductivity and topologically nontrivial bands is controllable by using proper tuning knobs, as detailed in the next section.

\section{Tunability}
\label{sec:tunable}

Various external tuning approaches have been extensively studied to modulate the physical properties of materials. The structural flexibility of 2D materials offers more opportunities for exploring the tuning impacts on their physical properties, including superconductivity, nontrivial band topology, coexistence of the two, and ultimately topological superconductivity. In this section, we will briefly review some of the latest advances in property for topological superconductivity, covering strain, gating, and ferroelectricity as the tuning knobs.

\subsection{Strain}
\label{sec:strain}

Strain is an effective way to introduce superconductivity by altering their electronic properties via changes in lattice structures for 2D materials. For example, $T_c \sim 3$ K was realized in a topological insulator of  Bi$_2$Te$_3$ between 3 to 6 GPa \cite{Zhang:2011aa}, and pressure-driven superconductivity and suppressed magnetoresistance were observed in WTe$_2$ \cite{Pan:2015aa,Kang:2015aa}. Furthermore, strain can induce a topological phase transition by narrowing the band gap associated with the strain-enhanced crystal field splitting for a system. When the band gap is decreased to a critical value or even closed, enabling the SOC to further hybridize the relevant bands and open an inverted gap at high-symmetry point(s) of BZ, a topological phase transition occurs. Through the band proximity effect mentioned before, the coexistence of superconductivity and nontrivial band topology renders the system to potentially become a TSC. 

One compelling example system is FeSe-based system. It has been shown that a monolayer FeSe exhibits surprisingly high  $T_c$ over $\sim 65$ K when grown on a STO substrate \cite{Wang:2012aa}. For such a system, it has been further proposed theoretically that the electronic structure can be tuned by the effective strain originating from a proper substrate, and develop emergent topological properties on top of its intrinsic superconducting property, pointing to the feasibility of realizing high-$T_c$ topological superconductors \cite{Hao:2014aa}. A few recent studies provided complementary experimental evidence for potential existence of topological superconductivity in monolayered FeSe systems \cite{Wang:2016aa,Shi:2017aa,Zhang:2017aa}. In one study, the robust electronic states along the edges of an FeSe monolayer on STO were identified by both STM and ARPES measurements, with the topological nature of the hosting FeSe nanoribbons confirmed by first-principles theory for a metastable state with chequerboard antiferromagnetic configuration \cite{Wang:2016aa}. In another, systematic spectroscopic indications support the occurrence of a topological quantum phase transition of FeTe$_{1-x}$Se$_x$/STO from the normal to topological state, induced by an increasing concentration of Te as a substitutional dopant to Se \cite{Shi:2017aa}. Here, aside from the stronger SOC effects associated with Te, the substitutional doping can also be viewed as applying a chemical strain, since Te possesses a larger atomic radius than Se. 

Beyond FeSe-based superconductors, impurity-assisted vortex states in LiFeAs have also been shown to exhibit MZMs \cite{Kong:2021aa}, attesting the nontrivial nature of the superconducting systems. Furthermore, as a compelling example for the strain-based tunability, a very recent study reported that the surface strain can alter the charge density waves (CDWs) appearing on the surfaces of the LiFeAs, and such CDWs can in turn regulate the special distribution of the vertex lattice that harbors the MZMs \cite{Li:2022aa}. More coverages on this line of advances can be found in a separate review within this volume \cite{Ding:2022aa}.

\begin{figure*}
\centering
\includegraphics[scale=0.5]{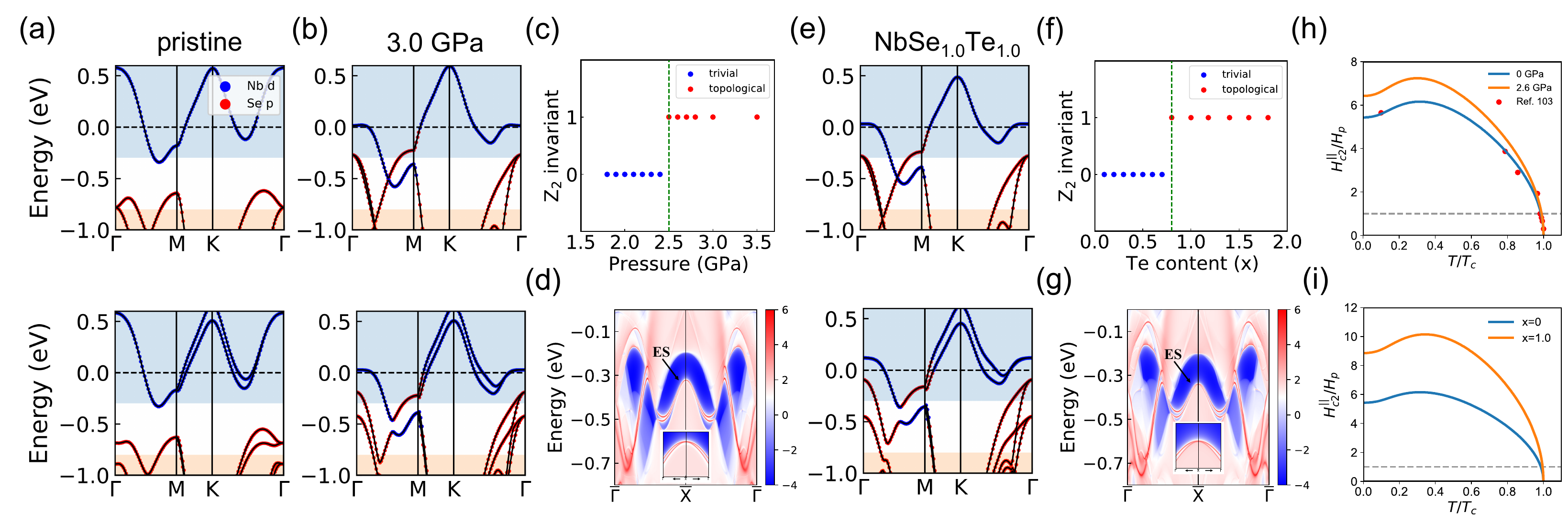}
\caption{(Color online) Band structures of 1$H$-NbSe$_2$ at (a) 0 GPa, (b) 3.0 GPa, and (e) Te doping concentration $x$ = 1.0 without (upper panel) and with (lower panel) the SOC. Calculated $Z_2$ invariants for 1$H$-NbSe$_2$ at different (c) pressures and (f) Te contents. Edge states (ES) of the semi-infinite slab along an Se-terminated Nb-zigzag edge at (d) 3.0 GPa and (g) $x$ = 1.0, with the Dirac nature at the $\mathrm{\bar{X}}$ point as highlighted in the inset. In-plane upper critical field $H^{||}_{c2}$ as a function of the transition temperature $T$ at (h) 2.6 GPa and (i) $x$ = 1.0. The red dots in (h) represent the experimental data extracted from \cite{Xi:2016aa}. Panels are modified from \cite{Li:2022ab} with permission from American Chemical Society.}
\label{fig:figure8}
\end{figure*}

Going even farther, beyond the Fe-based systems, a theoretical study has proposed that monolayer NbSe$_2$ can be converted into an Ising superconductor with nontrivial band topology via physical or chemical pressuring \cite{Li:2022ab}. The notion of Ising superconductivity was proposed recently in transition metal dichalcogenide (TMD) thin-film systems such as MoS$_2$ and NbSe$_2$ that possess strong SOC and inversion symmetry breaking \cite{Lu:2015aa,Xi:2016aa}. Specifically for monolayer NbSe$_2$, it was experimentally observed to exhibit Ising superconductivity \cite{Xi:2016aa}, signified by the surprisingly high in-plane upper critical field well above the Pauli paramagnetic limit \cite{Clogston:1962aa}. In such systems, the effective Zeeman field is generated by the broken inversion symmetry, which allows the inherently strong SOC of the systems to lock the spins of the electrons moving in-plane into the out-of-plane directions. When a strain was applied, a topological phase transition can be induced in monolayer NbSe$_2$, rendering the system a promising candidate to realize topological Ising superconductivity \cite{Li:2022ab}, as detailed below.

First, with increasing hydrostatic pressure, the Nb-$d$ and Se-$p$ bands of monolayer NbSe$_2$ within the 1$H$ phase (1$H$-NbSe$_2$) are moving closer towards each other, and then touch and cross, especially along the $\Gamma$-M path in the first BZ [see Figs.~\ref{fig:figure8}(a) and ~\ref{fig:figure8}(b)]. With the SOC included, the crossing bands open a gap, associated with an explicit band inversion around the M point [see Fig. ~\ref{fig:figure8}(b)]. The corresponding topological invariants $Z_2$ are summarized in Fig. ~\ref{fig:figure8}(c), confirming that the 1$H$-NbSe$_2$ is in the topologically nontrivial phase with $Z_2$ = 1 at 2.5 GPa or higher. Secondly, via substitutional doping of Se by Te, a chemical pressure was applied on 1$H$-NbSe$_2$, leading to similar band evolution [see Fig. ~\ref{fig:figure8}(e)]. As shown in Fig. ~\ref{fig:figure8}(f), when $x$ $\geq$ 0.8, the NbSe$_{2-x}$Te$_x$ system hosts a topologically nontrivial phase with $Z_2$ = 1. For both pressuring approaches, as another manifestation of the nontrivial topology in the band structures, the edge states of a semi-infinite slab along an Se-terminated Nb-zigzag edge of NbSe$_2$ at 3.0 GPa and NbSe$_{2-x}$Te$_x$ with $x$ = 1.0 are shown in Figs. ~\ref{fig:figure8}(d) and ~\ref{fig:figure8}(g), respectively. Finally, the $T_c$’s of 1$H$-NbSe$_2$ under different pressures and at the Te doping concentration of $x$ =1.0 were further estimated, showing relatively small variations from that of pristine 1$H$-NbSe$_2$. In particular, the Ising pairing nature has been explicitly enhanced, as indicated by the significantly enhanced in-plane upper critical fields [$H^{||}_{c2}$, see Figs. ~\ref{fig:figure8}(h) and ~\ref{fig:figure8}(i)], and also as confirmed by the preliminary experimental results of NbSe$_{2-x}$Te$_x$ flakes for a broad range of Te concentrations \cite{Li:2022ab}.

\subsection{Gating}
\label{sec:gating}

In condensed matter physics, gating has been a long and well-established approach to modulate the carrier densities and various corresponding physical properties. One major step forward surrounding this traditional approach was the recent developments of ionic liquid or solid gating, allowing to reach much higher carrier densities \cite{Ono:2009aa}, and enabling discoveries of emergent physical phenomena such as superconductivity in MoS$_2$ \cite{Ye:2012aa}. The enabling power of such an innovative gating approach has also been exemplified by the discovery of Ising superconductivity in MoS$_2$ \cite{Lu:2015aa}, a subject briefly reviewed in the preceding subsection.

In Sec.~\ref{sec:strain}, we spent more discussions on how strain can be used as an effective tuning knob to induce phase transitions of superconducting systems into the topologically nontrivial regime. The application of external electric field as a form of gating has also been demonstrated to be able to induce topological phase transitions \cite{Kim:2012aa}, but here, gating is mainly invoked as a means to convert topologically nontrivial and yet non-superconducting systems into the superconducting regime. One representative line of studies was about layered TMD systems. It was first predicted that such systems including MoS$_2$ and WTe$_2$ can be quantum spin Hall insulators under proper conditions or in proper structural phases \cite{Qian:2014aa}, and robust edge states suggestive of the nontrivial topology have been observed experimentally for WTe$_2$ \cite{Fei:2017aa,Tang:2017aa}. On the other hand, as insulators, such systems are non-superconducting. To induce superconductivity, proper charge doping into the systems is indispensable, as successfully demonstrated for WTe$_2$ via gating \cite{Sajadi:2018aa,Fatemi:2018aa}. It should be noted that, even though coexistence of nontrivial topology and superconductivity remains to be achieved, the fact that the same material platform of WTe$_2$ encompasses both essential properties of topological superconductivity offers new opportunities for further investigations. In this regard, it is also worth noting that a WTe$_2$ monolayer stabilized on an $s$-wave superconductor NbSe$_2$ can exhibit superconductivity while with the robust edge states preserved \cite{Lupke:2020aa}, but the approach is conceptually similar to that of proximity-induced topological superconductivity reviewed earlier in Sec.~\ref{sec:real-proximity} (e.g., Bi$_2$Te$_3$/NbSe$_2$ \cite{Xu:2014aa,Xu:2015aa}), instead of effective gating as emphasized here. 

\subsection{Ferroelectricity}
\label{sec:ferroelectricity}

As a nonvolatile, reversible, and thus more desirable approach, ferroelectric effects can be also exploited to modulate the superconductivity of an overlayer, given that a ferroelectric material harbors switchable polarization upon application of a voltage pulse. For some 3D conventional superconductors, such ferroelectric tuning of superconductivity has been achieved by forming heterostructures with ferroelectric films, such as significant $T_c$ modulations in Pb(Zr$_x$Ti$_{1-x}$)O$_3$/GdBa$_2$Cu$_3$O$_{7-x}$ \cite{Ahn:1999aa} and BiFeO$_3$/YBa$_2$Cu$_3$O$_{7-x}$ \cite{Crassous:2011aa}, and a complete switching of a superconducting transition in Nb-doped SrTiO$_3$ with Pb(Zr, Ti)O$_3$ as the ferroelectric overlayer \cite{Takahashi:2006aa}. However, the study of ferroelectric-tuned superconductivity in 2D systems has been very rare so far. In fact, the concurrent discoveries of 2D ferroelectric materials \cite{Liu:2016aa,Chang:2016aa,Ding:2017aa,Xiao:2018aa,Xue:2018aa,Yuan:2019aa} and 2D superconducting materials \cite{Xi:2016aa,Xu:2015ab,Yoshida:2018aa,Song:2021ab} should offer unprecedented opportunities for exploration of ferroelectrically tuned superconductivity and related devices, especially given the atomically sharp interfacial qualities of such van der Waals heterostructures \cite{Li:2020ac,Zhou:2021ac}.

Recently, ferroelectric switching of topological states has also been proposed theoretically in the van der Waals heterostructures of 2D trivial semiconducting/insulating and ferroelectric materials, such as Bi(111) bilayer/In$_2$Se$_3$ \cite{Bai:2020aa}, $\beta$-phase antimonene/In$_2$Se$_3$ \cite{Zhang:2021aa}, CuI/In$_2$Se$_3$ \cite{Marrazzo:2022aa}, and In$_2$Te$_3$/In$_2$Se$_3$ \cite{Huang:2022aa}. In these systems, the opposite polarization states are associated with different topological phases. Such a topological switching results from two aspects: one is that the ferroelectric polarization can change band alignments, band hybridizations, and charge transfer between the ferroelectric and trivial materials in the heterostructures.  The other is that the strong SOC effect of the heavy elements in the trivial layers can inverse the bands to induce the nontrivial band topology only in one polarization. 

Given the strong couplings between topological/superconducting and ferroelectric states, it is highly feasible to simultaneously tune superconductivity and band topology in 2D heterostructures using nonvolatile ferroelectric control, eventually realizing topological superconductivity. A very recent theoretical study has targeted to achieve this goal, presenting simultaneously tunable $T_c$ and band topology in a heterobilayer of superconducting IrTe$_2$ and ferroelectric In$_2$Se$_3$ monolayers \cite{Chen:2022aa}. The $T_c$ of the heterobilayer is shown to depend on the In$_2$Se$_3$ polarization, with the higher $T_c$ attributed to enhanced interlayer electron-phonon coupling when the polarization is downward. Meanwhile, the band topology is also switched from trivial to nontrivial as the polarization is reversed from upward to downward. Such 2D superconductor/ferroelectric heterostructures with coexistence of superconductivity and nontrivial band topology provide highly appealing candidates for realizing topological superconductivity, and this reversible and nonvolatile approach also offers promising new opportunities for detecting, manipulating, and ultimately braiding Majorana fermions.

\section{Conclusions and perspectives}
\label{sec:conclusion}

In this review, we have attempted to summarize some of the main developments surrounding 2D crystalline superconductors that possess either intrinsic $p$-wave pairing or nontrivial band topology. The selection of the contents has been made with the excellent review of Iwasa and collaborators \cite{Saito:2016ab} as the starting place, with some of own findings subjectively and hopefully proportionally incorporated. 

We have first introduced a classification of the generic topological superconductivity reached through three different conceptual schemes: (i) real-space superconducting proximity effect-induced TSC; (ii) reciprocal-space superconducting proximity effect-induced TSC; and (iii) intrinsic TSC. Whereas the first scheme has so far been most extensively explored, the other two remain to be fully substantiated and developed, including their subtle yet intrinsic differences.

For intrinsic or $p$-wave superconductors, we have reviewed the four candidate systems, old and new, that have been explored in the field, including Sr$_2$RuO$_4$, UTe$_2$, and graphene-based systems. In particular, among the 2D systems, we have predicted a Pb$_3$Bi alloyed system properly stabilized on a Ge(111) substrate to be an appealing candidate for realizing intrinsic topological superconductivity. 

For superconductivity with coexisting nontrivial band topology, we have reviewed the developments surrounding TMD monolayered systems as well as some of the candidates we identified. Notably, CoX (X = As, Sb, Bi) systems have been predicted to be stable in monolayered structural form; these systems may not only serve as new platforms for realizing high-$T_c$ superconductivity on STO, but may also harbor nontrivial band topology and robust edge states. Furthermore, we have shown experimentally that few-layered stanene films grown on Bi(111) with strong SOC are superconducting and possess robust edge states that can be attributed to their nontrivial band topology. 

In the pursuit for topological superconductivity, one widely recognized crucial aspect is the tunability of such systems and properties. We have attempted to outline three of the tuning knobs in acquiring one or both of the essential ingredients of TSC, including strain, gating, and ferroelectricity. In particular, strain has been demonstrated very recently as an effective means to regulate the spatial distribution and mutual interaction of a MZM lattice \cite{Li:2022aa}. Furthermore, we optimistically expect that the reversible and nonvolatile ferroelectric tunability will play a major role in gaining precise control and manipulation of MZMs. Irrespective of which dominant tuning scheme or their combination, the very fact of 2D systems will always offer superior advantages.

A further concern is what properties a TSC should possess in order to be favorable for potential applications in fault-tolerant quantum information processing. The first important feature is tunability; as discussed in Sec.~\ref{sec:tunable}, the tunability plays a powerful role in achieving topological superconductivity and regulating the distribution of the MZMs as well as their mutual interactions. The second is to have large superconducting gaps and high transition temperatures, which make the TSC systems more robust against temperature fluctuations and enable relatively easy manipulations of the MZMs. The third is short coherence length, which ensures the MZMs to survive in high magnetic fields, reduces the probability of the MZMs being pinned by impurities, and avoids fusion of the MZMs due to interacting magnetic vortices. Collectively, such merits will help to broaden the space for manipulating MZM-based qubits.

At present, discoveries of new TSC systems and unambiguous identification of MZMs remain to be the forefront challenges and advancing direction of the field. Given this status quo, it is premature and somewhat impractical to elaborate excessively on MZM braiding and non-Abelian statistics. Nevertheless, as outlined in the preceding section, the various enabling and complementary tuning capabilities being developed in the field will undoubtedly optimize and even maximize our chances to achieve these earnestly sought objectives step by step, casting the first ray of sunlight in the dream era of topological quantum computing.

\begin{acknowledgments}
We thank many of our collaborators who have contributed to the main findings highlighted in this review, especially Dr. Jinfeng Jia, Dr. Changgan Zeng, Dr. Bing Xiang, and Dr. Chenxiao Zhao on the experimental side, and Dr. Wenjun Ding, Dr. Leiqiang Li, Dr. Jianyong Chen on the theory side. This work was supported in part by the National Natural Science Foundation of China (Grant No. 11634011 and No.11974323), the National Key R\&D Program of China (Grant No.2017YFA0303500), the Anhui Initiative in Quantum Information Technologies (Grant No. AHY170000), the Strategic Priority Research Program of Chinese Academy of Sciences (Grant No. XDB30000000), and the Innovation Program for Quantum Science and Technology (Grant No. 2021ZD0302800).
\end{acknowledgments}

\end{document}